# Three-integral oblate galaxy models

## F.H.A. Robijn and P.T. de Zeeuw
*Sterrewacht Leiden, PO box 9513, 2300 RA Leiden, The Netherlands*

**ABSTRACT**
A simple numerical scheme is presented for the construction of three-integral phase-space distribution functions for oblate galaxy models with a gravitational potential of Stäckel form, and an arbitrary axisymmetric luminous density distribution. The intrinsic velocity moments can be obtained simultaneously with little extra effort. The distribution of the inner and outer turning points of the short-axis tube orbits that are populated can be specified freely, and is chosen in advance. The entire distribution function is then derived from the density by an iterative scheme that starts from the explicitly known distribution function of the thin-orbit (maximum streaming) model, in which only the tubes with equal inner and outer turning points are populated. The versatility and limitations of this scheme are illustrated by the construction of a number of self-consistent three-integral flattened isochrone models of Kuzmin–Kutuzov type, and by investigation of special cases where the scheme is tractable analytically. This includes the behaviour of the distribution functions in the outer regions of the models. The scheme converges rapidly for models containing orbits with ratios of the outer to inner turning point as large as ten, and is particularly suited for the construction of tangentially anisotropic flattened models, self-consistent as well as non-consistent. The algorithm simplifies in the disk and spherical limit, and can be generalized to triaxial models.

## 1 INTRODUCTION

The observable properties of elliptical galaxies indicate that their internal dynamics is governed by three integrals of motion (Binney 1976, 1978). For oblate systems two of the three are known, the energy $E$ and the angular momentum component $L_z$ along the symmetry axis. An exact third integral $I_3$ exists only for special classes of potentials, but adequate approximations have been derived for moderately flattened axisymmetric models (e.g., Saaf 1968; Innanen & Papp 1977; Gerhard & Saha 1991).

The construction of the full class of dynamical models for elliptical galaxies is a major undertaking. Progress has been made recently on a number of fronts, in particular for oblate systems. Even though elliptical galaxies as a class have triaxial shapes, the majority may well be nearly oblate (Franx, Illingworth & de Zeeuw 1991), so that oblate models are useful. Various practical methods have been developed for the construction of the special model with phase-space distribution function $f = f(E, L_z)$ (Hunter & Qian 1993; Dehnen & Gerhard 1994; Magorrian 1995; Kuijken 1995; Qian et al. 1995).

An exact third integral is known explicitly for the class of flattened models with a potential of Stäckel form (Kuzmin 1956; de Zeeuw 1985, hereafter dZ), and some self-consistent three-integral dynamical models of this type have been constructed, e.g., by numerical methods (Bishop 1986; 1987) or by series expansions (Dejonghe & de Zeeuw 1988, hereafter DZ). The distribution function for the model with the maximum possible streaming motions can be found by a single quadrature over the density (Bishop 1987; de Zeeuw & Hunter 1990, hereafter ZH). In oblate Stäckel models all orbits are short-axis tubes, but only those with vanishing radial action — which lie on spheroidal shells — are populated in the maximum streaming model. They are often referred to as *thin* (tube) orbits, and the corresponding model is called the *thin-orbit model*. These flattened models connect the sphere made exclusively of circular orbits with the similar axisymmetric disk.

When no exact $I_3$ is known, dynamical models can be constructed by numerical methods (e.g., Richstone 1980, 1982, 1984; Levison & Richstone 1985a, b) or by use of an approximate integral (Petrou 1983a, b). This approach has been employed recently by Dehnen & Gerhard (1993), who constructed a large family of approximate three-integral distribution functions for a flattened isochrone model, and investigated the relation between the internal dynamics and the observable kinematics. Their method is applicable to a wide variety of mass models with realistic density profiles. The one application that has been published so far is for a mass model that is nearly identical to the Kuzmin–Kutuzov model. This has a Stäckel potential, and its exact third integral has been used to construct a number of distribution functions (DZ, ZH).

Little is known about the stability of flattened galaxy models. Some N-body simulations have been carried out

(Merritt 1987; Merritt & Stiavelli 1990), but the paucity of available distribution functions to set up the initial conditions is one of the main reasons for our lack of knowledge. We have shown recently (Robijn & de Zeeuw 1995) that the linear stability analysis pioneered by Kalnajs (1977) for axisymmetric disks, and subsequently used by e.g., Polyachenko & Shukhman (1981), Palmer & Papaloizou (1987), Weinberg (1989, 1991) and Saha (1991, 1992) to study spherical models, also can be carried out for oblate Stäckel models. One of the first applications is a study of the thin orbit models, which have been shown by N-body simulations to be liable to ring- and lopsided instabilities, depending on the flattening of the model. Based on studies of spheres and flat disks, we expect that an increase in the amount of radial support will stabilize the radially 'cold' thin-orbit models. In order to investigate this, we need distribution functions for models in which not only the thin short-axis tubes are populated, but also 'thick' tube orbits with a finite radial extent. It is those models that we construct here.

The thin-orbit model has a distribution function of the form $f = f_{\rm tsm}(J_\phi, J_\nu)\delta(J_\lambda)$, where $J_\lambda$ is the radial action, $J_\phi = L_z$ is the azimuthal action, and $J_\nu$ is the latitudinal action. In this paper we write the distribution function in the (general) form $f = f_{\rm gsm}(J_\phi, J_\nu)g(J_\lambda, J_\phi, J_\nu)$, where $g$ is a preassigned function, and we show how to find $f_{\rm gsm}$, consistent with a given axisymmetric density $\rho$ in an oblate Stäckel potential $V$, by an iterative method, starting with the thin-orbit function $f_{\rm tsm}$ as a first guess for $f_{\rm gsm}$. We will consider functions $g$ that are peaked in $J_\lambda$, so that the models will be fairly close to the thin-orbit model, and few iterations are needed. The stability analysis of these models will be discussed in a subsequent paper.

Our method of specifying part of the distribution function, and solving for the remainder, is not new, and was used for flat disks by Shu (1969). Bishop (1986) applied it to oblate Stäckel models, starting from a different initial guess. Gerhard (1991) and Dehnen & Gerhard (1993) have recently popularized this approach for spherical and oblate models.

In Section 2 we define our notation, and present the construction method. A detailed description is given in Section 3, where we also investigate what properties of the assigned function $g$ are important for convergence of the iterative scheme. We illustrate the method by constructing a number of self-consistent Kuzmin–Kutuzov models with thick tubes. In Section 4 we consider special and limiting cases for which the algorithm simplifies, and where further insight in the method can be gained by analytic means. Concluding remarks follow in Section 5.

## 2 OBLATE GALAXY MODELS

We first summarize the basic properties of oblate Stäckel models, present the fundamental integral equation for their phase-space distribution functions, and then outline an iterative scheme for its solution. Derivations and further information can be found in dZ85, DZ, and ZH.

### 2.1 Orbits and integrals of motion

The motion in an oblate galaxy with a gravitational potential of Stäckel form is best described in prolate spheroidal coordinates $(\lambda, \nu, \phi)$. These are related to standard cylindrical coordinates $(R, z, \phi)$ by

$$R^2 = \frac{(\lambda+\alpha)(\nu+\alpha)}{(\alpha-\gamma)}, \qquad z^2 = \frac{(\lambda+\gamma)(\nu+\gamma)}{(\gamma-\alpha)}, \qquad (2.1)$$

where $\alpha$ and $\gamma$ are constants and the coordinates $\nu$ and $\lambda$ lie in the range $-\gamma \le \nu \le -\alpha \le \lambda$. Surfaces of constant $\lambda$ are prolate spheroids, while those of constant $\nu$ are two-sheeted hyperboloids. The foci are located at $z = \pm\sqrt{\gamma-\alpha}$. Each set of $(\lambda, \nu, \phi)$ corresponds in general to two points $(R, \pm z, \phi)$. In these coordinates the potential $V(\lambda, \nu)$ takes the form:

$$V = -\frac{(\lambda+\gamma)G(\lambda) - (\nu+\gamma)G(\nu)}{(\lambda-\nu)}, \qquad (2.2)$$

where $G(\tau)$ is an arbitrary smooth function that determines the shape of the potential, and $\tau = \lambda, \nu$.

The equations of motion separate in the $(\lambda, \nu, \phi)$ coordinates. Since the potential is axisymmetric, the momentum $p_\phi$ conjugate to $\phi$ is constant, and equals $L_z = R^2\dot\phi$, the component of the angular momentum parallel to the $z$-axis. The motion in $\lambda$ and $\nu$, i.e., in the meridional plane, is described by

$$p_\tau^2 = \frac{B(\tau)}{2(\tau+\alpha)^2(\tau+\gamma)}, \qquad (\tau = \lambda, \nu), \qquad (2.3)$$

where

$$B(\tau) = (\tau+\alpha)(\tau+\gamma)E - (\tau+\gamma)I_2 - (\tau+\alpha)I_3 - U(\tau). \quad (2.4)$$

and

$$U(\tau) = -(\tau+\alpha)(\tau+\gamma)G(\tau). \qquad (2.5)$$

Here $E$ is the total orbital energy, $I_2 = \tfrac{1}{2}L_z^2$, and $I_3$ is the third isolating integral of motion given by (cf. eq. [2.13] of DZ)

$$I_3 = \tfrac{1}{2}(L_x^2 + L_y^2) + (\gamma-\alpha)\left[\tfrac{1}{2}v_z^2 - z^2\frac{G(\lambda)-G(\nu)}{\lambda-\nu}\right]. \qquad (2.6)$$

Each set of values of $E$, $I_2 \ge 0$ and $I_3$ for which $p_\lambda^2 \ge 0$ and $p_\nu^2 \ge 0$ in some range of $\lambda$ and $\nu$, respectively, corresponds to an orbit. It is bound when $E \le 0$. In this case the function $B(\tau)$ generally has three zeroes for $\tau = \nu_0, \lambda_1, \lambda_2$, and each orbit fills an area in the meridional plane defined by

$$-\gamma \le \nu \le \nu_0, \qquad \lambda_1 \le \lambda \le \lambda_2. \qquad (2.7)$$

Orbits of this shape are usually referred to as *short-axis tubes*.

The constants $\nu_0, \lambda_1, \lambda_2$ are functions of $E, I_2, I_3$, and are called the *turning points* of the orbit. ZH have shown that the relations between the standard integrals $(E, I_2, I_3)$ and $(\nu_0, \lambda_1, \lambda_2)$ can be written as

$$\begin{aligned}
E &= U[\nu_0, \lambda_1, \lambda_2], \\
I_2 &= \frac{(-\alpha-\nu_0)(\lambda_1+\alpha)(\lambda_2+\alpha)}{\gamma-\alpha}U[-\alpha, \nu_0, \lambda_1, \lambda_2], \\
I_3 &= \frac{(\nu_0+\gamma)(\lambda_1+\gamma)(\lambda_2+\gamma)}{\gamma-\alpha}U[-\gamma, \nu_0, \lambda_1, \lambda_2],
\end{aligned} \qquad (2.8)$$

where the square brackets indicate *divided differences* of the function $U(\tau)$ defined in equation (2.5). These are defined iteratively by

$$U[\tau_1, \tau_2] = \frac{U(\tau_1) - U(\tau_2)}{\tau_1 - \tau_2}, \qquad (2.9a)$$

**Figure 1.** The three-dimensional volume filled by a short-axis tube orbit in an oblate Stäckel model. The thin solid lines are the intersections of the prolate spheroidal coordinates $(\lambda, \nu, \phi)$ in which the motion separates with the equatorial plane $z = 0$ and with two meridional planes $(R, z)$ at $\phi = 0$ and $\phi = \pi/2$. The dot indicates the location of the focus along the positive $z$-axis. The orbital volume is bounded by four prolate spheroidal coordinate surfaces: the top and bottom surfaces are parts of hyperboloids of revolution, labelled by the turning point $\nu_0$, while the inner and outer boundaries are spheroids of revolution labelled by the turning points $\lambda_1$ and $\lambda_2$.

and

$$U[\tau_1, \tau_2, ..., \tau_n] = \frac{U[\tau_1, \tau_3, ..., \tau_n] - U[\tau_2, \tau_3, ..., \tau_n]}{\tau_1 - \tau_2}. \quad (2.9b)$$

The ordering of the arguments is not significant. With this notation the function $B(\tau)$ of equation (2.4) becomes

$$B(\tau; \nu_0, \lambda_1, \lambda_2) = (\tau - \nu_0)(\tau - \lambda_1)(\lambda_2 - \tau)U[\tau, \nu_0, \lambda_1, \lambda_2]. \quad (2.10)$$

Centrally concentrated models have $U'''(\tau) > 0$ for $\tau \geq -\gamma$. This guarantees that all third order divided differences $U[\tau_1, \tau_2, \tau_3, \tau_4]$ are strictly positive (Hunter & de Zeeuw 1992), so that $B(\tau)$ has no more than three zeroes: all orbits are short-axis tubes.

The three *action integrals* $J_\tau = (2\pi)^{-1} \oint p_\tau d\tau$ can be written as follows:

$$J_\lambda = \frac{\sqrt{2}}{\pi} \int_{\lambda_1}^{\lambda_2} \sqrt{\frac{B(\lambda; \nu_0, \lambda_1, \lambda_2)}{\lambda + \gamma}} \frac{d\lambda}{\lambda + \alpha},$$

$$J_\phi = L_z = \sqrt{\frac{-2B(-\alpha; \nu_0, \lambda_1, \lambda_2)}{\gamma - \alpha}}, \quad (2.11)$$

$$J_\nu = \frac{\sqrt{2}}{\pi} \int_{-\gamma}^{\nu_0} \sqrt{\frac{B(\nu; \nu_0, \lambda_1, \lambda_2)}{\nu + \gamma}} \frac{d\nu}{(-\alpha - \nu)}.$$

The integrals for $J_\lambda$ and $J_\nu$ generally need to be evaluated numerically.

### 2.2 Distribution functions

The fundamental integral equation for the phase-space distribution function $f_{sm}(\lambda, \nu, v_\lambda, v_\phi, v_\nu)$ that gives rise to a density $\rho_m(\lambda, \nu)$ in a gravitational potential $V(\lambda, \nu)$ is

$$\rho_m(\lambda, \nu) = \iiint f_{sm}(\lambda, \nu, v_\lambda, v_\phi, v_\nu) \, dv_\lambda dv_\phi dv_\nu. \quad (2.12)$$

Because $V$ is here of Stäckel form, each orbit has three exact isolating integrals of motion, so that Jeans' theorem is valid: $f_{sm}$ is a function of the three integrals of motion, so we can consider $f_{sm} = f_{sm}(E, I_2, I_3)$, or $f_{sm} = f_{sm}(\nu_0, \lambda_1, \lambda_2)$, or $f_{sm} = f_{sm}(J_\lambda, J_\phi, J_\nu)$. In each case, transformation of $dv_\lambda dv_\phi dv_\nu$ to $dEdI_2dI_3$ etc., with the appropriate Jacobian determinant, gives the relevant form of the fundamental integral equation (2.12). DZ (eq. [3.2]) write (2.12) in terms of $(E, I_2, I_3)$, while ZH discuss its forms in terms of the turning points $(\nu_0, \lambda_1, \lambda_2)$ and the actions $(J_\lambda, J_\phi, J_\nu)$ (their eqs. [2.23] and [2.47]).

Since $f_{sm}$ is a function of three arguments, and $\rho_m$ depends on only two variables, many different $f_{sm}$'s will be consistent with the same $\rho_m$, so that equation (2.12) has many solutions. Two of these are readily available. The first is the special model with $f_{sm} = f_{sm}(E, L_z)$, in which the orbits are populated such that there is no net dependence on the third integral. Its distribution function can be found by application of the Hunter & Qian (1993) method, which requires (numerical) evaluation of a contour integral. The second is the so-called thin-orbit model, in which the stars occupy only the short axis tubes that have no $\lambda$-excursion, and hence lie on prolate spheroidal shells. In this case $f_{sm} = f_{tsm}(J_\phi, J_\nu)\delta(J_\lambda)$. Bishop (1987) and ZH have shown that $f_{tsm}$ can be found by a single real quadrature.

We are interested in distribution functions that populate not only the thin orbits, but also those with a finite $\lambda$-extent. Instead of the turning points $\lambda_1$ and $\lambda_2$, we employ the quantities

$$\lambda_m = \tfrac{1}{2}(\lambda_1 + \lambda_2), \qquad \epsilon = \tfrac{1}{2}(\lambda_2 - \lambda_1). \quad (2.13)$$

Here $\epsilon \geq 0$ controls the 'thickness' of the short-axis tube, and $\lambda_m$ indicates its mean location in the radial direction (Figure 1). When $\epsilon = 0$ the two radial turning points $\lambda_1$ and $\lambda_2$ coincide, so that the 'radial' action $J_\lambda = 0$, and the orbit is a thin short-axis tube. The relations between the standard integrals $(E, I_2, I_3)$ and $(\nu_0, \lambda_m, \epsilon)$ follow from equation (2.8), upon substitution of $\lambda_1 = \lambda_m - \epsilon$, $\lambda_2 = \lambda_m + \epsilon$.

With the definitions (2.13), equation (2.23) of ZH can be transformed to the fundamental integral equation in terms of the three integrals $\nu_0, \lambda_m, \epsilon$:

$$\rho_m(\lambda, \nu) =$$

$$4\sqrt{2} \int_\nu^{-\alpha} d\nu_0 \int_{\frac{1}{2}(\lambda+\alpha)}^{\infty} d\lambda_m \int_{|\lambda-\lambda_m|}^{\lambda_m+\alpha} d\epsilon \, \frac{U^*(\lambda, \nu; \nu_0, \lambda_1, \lambda_2)}{\sqrt{(\nu_0-\nu)(-\alpha-\nu_0)(\lambda-\nu_0)}} \quad (2.14)$$

$$\times \frac{\epsilon[(\lambda_m-\nu_0)^2 - \epsilon^2] f_{sm}(\nu_0, \lambda_m, \epsilon)}{\sqrt{[(\lambda_m-\nu)^2-\epsilon^2][(\lambda_m+\alpha)^2-\epsilon^2][\epsilon^2-(\lambda-\lambda_m)^2]}},$$

where

$$U^* = \frac{U[\nu_0, \lambda_1, \lambda_1, \lambda_2] U[\nu_0, \lambda_1, \lambda_2, \lambda_2] U[\nu_0, \nu_0, \lambda_1, \lambda_2]}{\sqrt{U[\nu_0, \lambda_1, \lambda, \lambda_2] U[\nu, \nu_0, \lambda_1, \lambda_2] U[\nu_0, -\alpha, \lambda_1, \lambda_2]}} \quad (2.15)$$

and we still have to substitute $\lambda_1 = \lambda_m - \epsilon$, $\lambda_2 = \lambda_m + \epsilon$. The area of integration in the $(\lambda_m, \epsilon)$ plane is illustrated in Figure 2a.

### 2.3 Iterative scheme

Our aim is to construct distribution functions $f_{sm}(\nu_0, \lambda_m, \epsilon)$ that also populate orbits with non-zero thickness $\epsilon > 0$.

**Figure 2.** Integration areas for the fundamental integral equation. a) In the $(\epsilon, \lambda_m)$-plane, defined in equation (2.13). b) In the $(s, t)$-plane, defined in equations (3.2) and (3.9). The light shaded regions indicate the full integration areas: all orbits with inner and outer $\lambda$-turning points that correspond to values of $(\epsilon, \lambda_m)$ or $(s, t)$ in these areas contribute density on the spheroidal shell with coordinate $\lambda$ in configuration space. The specific choice (3.6) for the function $g_{sm}$ only populates orbits up to a maximum relative thickness $s_{\max}$, so that the integration over $s$ runs between 0 and $s_{\max}$, as indicated by the dark shaded regions. The thin-orbit model has $s_{\max} = 0$, so that the integration areas shrink to a point, indicated by the filled squares.

In the spirit of Bishop (1986), we resolve the distribution function $f_{sm}$ as the following product

$$f_{sm}(\nu_0, \lambda_m, \epsilon) = f_{gsm}(\nu_0, \lambda_m)\tilde{g}_{sm}(\nu_0, \lambda_m, \epsilon), \qquad (2.16)$$

where $\tilde{g}_{sm}$ gives the distribution of the radial excursions $\epsilon$ of the orbits, which may depend on the values of the latitudinal turning points $\nu_0$ and the mean radial positions $\lambda_m$. Writing $f_{sm}$ in this way does not imply any restrictions; all distribution functions can be split up as in (2.16). If we specify $\tilde{g}_{sm}$, and substitute the result in equation (2.14), we are left with an integral equation for $f_{gsm}(\nu_0, \lambda_m)$. The choice of $\tilde{g}_{sm}$ determines what function will be found.

We choose to normalize the function $\tilde{g}_{sm}$ such that

$$\int_0^\infty \tilde{g}_{sm}(\nu_0, \lambda_m, \epsilon)\, dJ_\lambda = 1, \qquad (2.17)$$

for fixed $(\nu_0, \lambda_m)$. It then follows that when $\tilde{g}_{sm} = \delta(J_\lambda)$, the function $f_{gsm}$ is identical to $f_{tsm}$, the thin orbit distribution function of Bishop (1987) and ZH. It is given by

$$\begin{aligned}f_{tsm}(\lambda_m, \nu_0) =& \frac{1}{8\pi^2\sqrt{\lambda_m + \gamma}(\lambda_m - \nu_0)U[\nu_0, \lambda_m, \lambda_m, \lambda_m]} \\ &\times \Big[(\lambda_m + \alpha)\rho(\lambda_m, -\alpha) - \sqrt{(-\alpha - \nu_0)}U[\nu_0, -\alpha, \lambda_m, \lambda_m] \\ &\times \int_{\nu_0}^{-\alpha} \frac{[\partial(\lambda_m - \sigma)\rho(\lambda_m, \sigma)/\partial\sigma]d\sigma}{\sqrt{(\sigma - \nu_0)}U[\sigma, \nu_0, \lambda_m, \lambda_m]}\Big],\end{aligned} \qquad (2.18)$$

where $\rho$ equals $\rho_m$, the model density. The divided differences of $U$ are always derived from the model potential $V$, but the above expression gives $f_{tsm}$ for any axisymmetric density $\rho$ in the potential $V(\lambda, \nu)$.

When the function $\tilde{g}_{sm}$ is sharply peaked near $\epsilon = 0$ (i.e., near $J_\lambda = 0$), the solution $f_{gsm}$ of (2.14) should be very similar to the thin orbit function $f_{tsm}$. This suggests the following approach. We start with the zeroth-order distribution function

$$f_0(\nu_0, \lambda_m, \epsilon) = f_{tsm}\{\rho_m\}\tilde{g}_{sm}(\nu_0, \lambda_m, \epsilon), \qquad (2.19)$$

where $f_{tsm}\{\rho_m\}$ is short-hand for the thin-orbit function $f_{tsm}(\lambda_m, \nu_0)$ that follows from equation (2.18) upon substitution of $\rho = \rho_m$. Since $\tilde{g}_{sm}$ is not the delta function in $J_\lambda$ that is appropriate for $f_{tsm}\{\rho_m\}$, the residual density

$$\rho_1 = \rho_m - \iiint f_{tsm}\{\rho_m\}\tilde{g}_{sm}(\nu_0, \lambda_m, \epsilon)\, d^3v, \qquad (2.20)$$

does not vanish everywhere, although we expect it to be much smaller than $\rho_m$. We have $f_{gsm} = f_0 + f_c$, where $f_c$ is the solution of

$$\rho_1 = \iiint f_c(\nu_0, \lambda_m)\tilde{g}_{sm}(\nu_0, \lambda_m, \epsilon)\, d^3v. \qquad (2.21)$$

This is the same integral equation as (2.14), but now for the density $\rho_1$ in the potential $V(\lambda, \nu)$. For sharply-peaked $\tilde{g}_{sm}$ we approximate $f_c$ by $f_1 = f_{tsm}\{\rho_1\}$. Taking as first order approximation $f_{gsm} = f_{tsm}\{\rho_m\} + f_{tsm}\{\rho_1\}$ then leads to a residual density $\rho_2$, which should be smaller than $\rho_1$. We can repeat this process as many times as we want, which leads to the following algorithm:

$$f_{gsm}(\nu_0, \lambda_m) = \sum_{i=0}^n f_{tsm}\{\rho_i\}, \qquad (2.22)$$

where

$$\begin{aligned}\rho_0 &= \rho_m(\lambda, \nu), \\ \rho_{i+1} &= \rho_i - \iiint f_{tsm}\{\rho_i\}\tilde{g}_{sm}\, d^3v, \quad (i = 1, \ldots, n).\end{aligned} \qquad (2.23)$$

If the residual densities $\rho_i$ decrease with increasing $i$, the series (2.22) will provide an increasingly better approximation to the actual distribution function. If this process converges, we still have to check that the resulting $f_{gsm}$ is non-negative everywhere. If it is not, it is not a physical distribution function, and another choice needs to be made for $\tilde{g}_{sm}$.

When $\tilde{g}_{sm}$ is sharply peaked, the zeroth order approximation (2.19) may already be adequate. Shu (1969) used it to construct self-consistent flat circular disks with nearly circular orbits. For less sharply peaked functions the iterative scheme (2.22)–(2.23) should work very well. However, as the thickness of the populated orbits increases, the density residuals $\rho_i$ will become larger and it is not clear *a priori* whether the algorithm will converge. We have implemented the algorithm, and have constructed a number of models. It turns out that convergence is reached easily for models with quite 'fat' orbits $\epsilon \sim 0.7(\lambda_m + \alpha)$, but that the number of required iterations increases strongly for broad $\tilde{g}_{sm}$ functions. The algorithm is described in detail in Section 3.

### 2.4 Kuzmin-Kutuzov mass models

We illustrate our method by applying it to the construction of self-consistent models with potential

$$V(\lambda, \nu) = -\frac{GM}{\sqrt{\lambda} + \sqrt{\nu}}, \quad (2.24)$$

and associated density

$$\rho_m(\lambda, \nu) = \frac{M\gamma}{4\pi} \frac{\alpha(\lambda + 3\sqrt{\lambda\nu} + \nu) - \lambda\nu}{(\lambda\nu)^{3/2}(\sqrt{\lambda} + \sqrt{\nu})^3}. \quad (2.25)$$

This mass model was introduced by Kuzmin (1956) and connects Kuzmin's (1953) flat circular disk ($\gamma = 0$) with Hénon's (1959) spherical isochrone ($\gamma = \alpha$). The surfaces of constant density are smooth and nearly oblate spheroidal with an axis ratio $\sim \sqrt{\gamma/\alpha}$, and become slightly less flattened at large radii. Kuzmin & Kutuzov (1962) showed that the distribution function $f_{sm}(E, L_z)$ could be found as a series expansion in powers of $E$ and $L_z$. Many properties of these Kuzmin–Kutuzov models were described by DZ, who also derived a closed form for $f_{sm}(E, L_z)$, albeit with a typographical error (see Batsleer & Dejonghe 1993). ZH showed that the thin-orbit distribution function $f_{sm} = f_{tsm}(\nu_0, \lambda_m)\delta(J_\lambda)$ (Figure 3) can be given in terms of elementary functions, and discussed its properties in detail.

## 3 DESCRIPTION OF THE METHOD

We first discuss a practical way to choose the function $\tilde{g}_{sm}$, introduce more convenient variables, and show that the convergence of the iterative scheme depends mostly on the moments of $\tilde{g}_{sm}$. Then we show how kinematic properties of the models can be calculated with little extra effort, and we briefly describe the numerical implementation.

### 3.1 Normalization of $\tilde{g}_{sm}$: the function $g_{sm}$

The normalization of $\tilde{g}_{sm}$ is to some extent arbitrary, as is the factorization (2.16) of the distribution function. All we require is that $\tilde{g}_{sm}$ is normalized such that in the limit of thin orbits only, we recover $f_{tsm}$. The most natural normalization is to take the condition (2.17), but at constant values of the actions $(J_\phi, J_\nu)$ rather than at constant values of the integrals $(\nu_0, \lambda_m)$. However, the transformation $(\nu_0, \lambda_m, \epsilon) \leftrightarrow (J_\lambda, J_\phi, J_\nu)$ generally requires numerical inversion, so that in fact both these normalizations are not very practical. We therefore work with a function $g_{sm}(\nu_0, \lambda_m, s)$ defined by

$$\tilde{g}_{sm}(\nu_0, \lambda_m, \epsilon) = \frac{c_g(\nu_0, \lambda_m)}{(\lambda_m + \alpha)\sqrt{\lambda_m - \nu_0}} g_{sm}(\nu_0, \lambda_m, s), \quad (3.1)$$

where

$$s = \frac{\epsilon}{\lambda_m + \alpha}, \quad (3.2)$$

so that $s$ is an integral of motion which gives the *relative thickness* of the orbit, and $0 \leq s \leq 1$. We require that

$$\int_0^1 g_{sm}(\nu_0, \lambda_m, s) \, ds^2 = 1, \quad (3.3)$$

at fixed $\nu_0$ and $\lambda_m$. By comparison of equations (3.3) and (2.17) it then follows that

$$c_g(\nu_0, \lambda_m) = \frac{(\lambda_m + \alpha)\sqrt{\lambda_m - \nu_0}}{\int_0^1 ds^2 \, g_{sm}(\nu_0, \lambda_m, s) |\partial J_\lambda/\partial s^2|_{(\nu_0, \lambda_m)}}, \quad (3.4)$$

which therefore depends on the choice of $g_{sm}$. The partial derivative of $J_\lambda$ can be written as a single quadrature, and is given in Appendix A. It is even in $s$, and hence a function of $s^2$.

We have not incorporated the factor $(\lambda_m + \alpha)\sqrt{\lambda_m - \nu_0}$ in the definition of $c_g$, because it diverges on orbits with $\lambda_m = \nu_0 = -\alpha$, i.e., on the oscillations along the $z$-axis that just reach the foci of the spheroidal coordinates. We will come back to the behaviour of the algorithm near the foci in Section 4.2. In the limit $\tilde{g}_{sm} = \delta(J_\lambda)$, so that only thin orbits are populated, we have $s = 0$, $g_{sm}(s) = \delta(s^2)$, and

$$c_g(\nu_0, \lambda_m) = \frac{\sqrt{2(\lambda_m + \gamma)}}{\sqrt{U[\nu_0, \lambda_m, \lambda_m, \lambda_m]}}. \quad (3.5)$$

This is well-behaved for all physical values of $-\gamma \leq \nu_0 \leq -\alpha \leq \lambda_m$.

### 3.2 Choice of $g_{sm}$; moments

Any distribution function $f_{sm}$ can be written as $f_{gsm} c_g g_{sm}$. As an example, Figure 4 shows both the factor $f_{gsm} c_g$ and the factor $g_{sm}$ for the two-integral Kuzmin–Kutuzov model with $f_{sm}(E, L_z)$. Both factors vary smoothly, but the function $g_{sm}$ shows a range of behaviour as a function of $s^2$, depending on the values of $\lambda_m, \nu_0$. In this paper we consider functions $g_{sm}$ that are even in $s$, and are of the form

$$g_{sm}(s) = \begin{cases} \frac{q+1}{s_{max}^2} \left(1 - \frac{s^2}{s_{max}^2}\right)^q & \text{for } 0 \leq s \leq s_{max}, \\ 0 & \text{for } s \geq s_{max}, \end{cases} \quad (3.6)$$

with $q > -1$ and $0 \leq s_{max} \leq 1$. These functions are all normalized as in equation (3.3), and show a similar range of shapes as seen in Figure 4. In principle, we can choose the maximum relative thickness $s_{max}$ to be a function of $\nu_0$ and $\lambda_m$, but we do not do so here, and from now on we suppress the dependence of $g_{sm}$ on $\nu_0$ and $\lambda_m$ in the expressions that follow. Figure 5 illustrates the cases $q = 0, 1, 2$.

We define the moments $\langle s^{2n} g_{sm} \rangle$ of $g_{sm}$ by

$$\langle s^{2n} g_{sm} \rangle = \frac{1}{n!} \int_0^1 s^{2n} g_{sm}(s) \, ds^2, \quad (3.7)$$

**Figure 3.** The distribution function $f_{\text{tsm}}(\lambda_m, \nu_0)$ for an E5 Kuzmin–Kutuzov model with $\alpha = -1$ and $\gamma = -0.25$. The thick solid curves are contours spaced logarithmically at intervals of 2. The focal corner lies at $\lambda_m = \nu_0 = 1$. See also Figure 2 of ZH.

**Figure 4.** The distribution function $f_{\text{sm}}(E, L_z^2)$ for an E5 Kuzmin–Kutuzov model with $\alpha = -4/9$ and $\gamma = -1/9$. The surface is the $f_{\text{gsm}}$ part, which includes the normalizations factor $c_g$. The thick solid curves are contours spaced logarithmically at intervals of 2. The focal corner lies at $\lambda_m = \nu_0 = 4/9$. The small diagrams in the upper half show the behaviour of $g_{\text{sm}}$ as a function of $s$, on the interval $[0, 1]$.

for $n = 0, 1, \ldots$. It follows that $\langle g_{\text{sm}} \rangle = \langle s^0 g_{\text{sm}} \rangle = 1$, by the normalization (3.3). The higher moments can be given explicitly for the choice (3.6):

$$\langle s^{2n} g_{\text{sm}} \rangle = \frac{\Gamma(q+2)}{\Gamma(n+q+2)} s_{\max}^{2n}. \tag{3.8}$$

When $s_{\max} \to 0$, we have $g_{\text{sm}}(s) = \delta(s^2)$, and all higher order moments vanish. When $0 < s_{\max} \ll 1$, the higher moments decrease very rapidly with increasing $n$.

**Figure 5.** Three different functions $g_{\rm sm}$ of the form (3.6) with $q = 0$ (solid line), $q = 1$ (dotted line), $q = 2$ (dashed line). They are normalized with respect to $s^2$ (see eq. [3.3]).

### 3.3 New variables

The fundamental integral equation appears in our iterative scheme in the form (2.23). We transform it to a more useful form by means of an alternative set of variables. In addition to the relative orbital thickness $s$, defined in equation (3.2), we introduce

$$t = \frac{\lambda - \lambda_m}{\lambda_m + \alpha}, \qquad u = \frac{-\alpha - \nu_0}{-\alpha - \nu}, \qquad x = \frac{-\alpha - \nu}{\lambda - \nu}, \qquad (3.9)$$

so that $0 \le t$, $0 \le u$, and $0 \le x \le 1$. These definitions mix the turning point variables $(\nu_0, \lambda_m)$ with the coordinates $(\lambda, \nu)$, but they facilitate the analysis of the fundamental integral equation. Carrying out the substitutions (3.9), and rearranging various terms results in

$$\rho_{i+1} = \rho_i - \int_0^1 du\, w_1(u) \int_0^{s_{\max}^2} ds^2 g_{\rm sm}(s) \times \\ \int_{-s}^{s} dt\, \frac{w_2(s,t,u)}{\sqrt{s^2 - t^2}} f_{\rm tsm}\{\rho_i\}, \qquad (3.10)$$

where we have written

$$w_1 = \frac{4\sqrt{2}}{\sqrt{u(1-u)(1-x+xu)}},$$
$$w_2 = \frac{[[1-x+xu(1+t)]^2 - (1-x)^2 s^2]}{\sqrt{(1+tx)^2 - (1-x)^2 s^2}\sqrt{1-x+xu(1+t)}} \qquad (3.11)$$
$$\times \frac{U^*(s,t,u)c_g(t,u)}{(1+t)^{3/2}\sqrt{1-s^2}},$$

and $c_g$ and $U^*$ are defined in equations (3.4) and (2.15), respectively. The thin orbit function $f_{\rm tsm}\{\rho_i\}$ is defined in equation (2.18), and is independent of $s$. Both $w_1$ and $w_2$ depend also on the coordinates $\lambda$ and $\nu$, but we suppress them as arguments because they are not integration variables. The square root of $s^2 - t^2$ vanishes at $s = t = 0$, which lies in the area of integration (Figure 2b). For this reason we have not incorporated it in the definition of $w_2$. Finally we remark that $w_2$ is an even function of $s$, and hence depends on $s^2$.

### 3.4 Convergence

When $s_{\max} \ll 1$, the function $g_{\rm sm}$ is sharply peaked, and it is useful to expand it in a series of derivatives of delta functions (e.g., Fridman & Polyachenko 1984, p. 150):

$$g_{\rm sm}(s) = \sum_{n=0}^{\infty} (-1)^n \langle s^{2n} g_{\rm sm} \rangle \delta^{(n)}(s^2), \qquad (3.12)$$

where the $\langle s^{2n} g_{\rm sm} \rangle \propto s_{\max}^{2n}$ are the moments of $g_{\rm sm}$ defined in equation (3.7), and $\delta^{(n)}$ is the $n^{\rm th}$ derivative of the delta function, which satisfies the relation

$$\int_{-}^{+} \delta^{(n)}(x) h(x)\, dx = (-1)^n h^{(n)}(0). \qquad (3.13)$$

We use this expansion to show that the convergence of our iterative scheme depends mostly on the moments of $g_{\rm sm}$, and less on its detailed functional behaviour.

Substitution of (3.12) in equation (3.10) gives

$$\rho_{i+1} = \rho_i - \int_0^1 du\, w_1(u) \sum_{n=0}^{\infty} \langle s^{2n} g_{\rm sm} \rangle \frac{d^n W_2(0,u)}{d(s^2)^n}, \qquad (3.14)$$

where we have written the $t$-integral as

$$W_2(s,u) = \int_{-s}^{s} dt\, \frac{w_2(s,t,u)}{\sqrt{s^2 - t^2}} f_{\rm tsm}\{\rho_i\}, \qquad (3.15)$$

and we have used the definition (3.13) to carry out the $s$-integration. The first term in the series expansion for $g_{\rm sm}$ is $\delta(s^2)$. Its contribution to the right hand side of equation (3.14) equals $-\rho_i$. Upon substitution of the specific form (3.8) for the moments, we are therefore left with

$$\rho_{i+1} = -\sum_{n=1}^{\infty} \frac{\Gamma(q+2)}{\Gamma(n+q+2)} s_{\max}^{2n} \int_0^1 du\, w_1(u) \frac{d^n W_2(0,u)}{d(s^2)^n}. \quad (3.16)$$

This is valid for all functions $g_{\rm sm}$ of the form (3.6) with $s_{\max} < 1$.

Since $0 \le t \le s \le s_{\max}$, it follows that both $s$ and $t$ are small when $g_{\rm sm}$ is sharply peaked. The functions $w_2(s,t,u)$ and $f_{\rm tsm}\{\rho_i\}$ then vary little over the integration area, and so does $W_2(s,u)$. Its derivatives with respect to $s^2$ are finite at $s = 0$. Since $W_2$ contains the thin orbit function $f_{\rm tsm}\{\rho_i\}$ as a factor, and since this is independent of $s$ and proportional to $\rho_i$, it follows that all the derivatives of $W_2$ are similarly proportional to $\rho_i$.

For small $s_{\max}$, the first term in the series on the right hand side of equation (3.16) dominates. This means that the residual density $\rho_{i+1}$ is roughly proportional to the residual $\rho_i$ times the first moment of the function $g_{\rm sm}$. Since this is proportional to $s_{\max}^2$, this shows why even for moderately peaked functions our iterative scheme converges rapidly. It furthermore shows that the shape of $g_{\rm sm}$ is less important than its moments. We can vary the shape of $g_{\rm sm}$ without significantly affecting the residual density, as long as we do not dramatically change the lowest order moments of $g_{\rm sm}$. Since $f_{\rm tsm}\{\rho_i\}$ is derived from these residuals, the solutions $f_{\rm gsm}(\lambda_m, \nu_0)$ that correspond to different $g_{\rm sm}$ will be rather similar as long as the first moments of these $g_{\rm sm}$ functions

**Figure 6.** Successive steps of the iterative scheme to obtain $f_{\rm gsm}$ for an E5 Kuzmin–Kutuzov model with $\alpha = -1, \gamma = -0.25$, and a function $g_{\rm sm}$ with parameters $q = 2$ and $s_{\rm max}^2 = 0.9$. Shown is the residual density $\rho_i/\rho_{\rm m}$ for each step. The bottom right plot shows the ratio of the resulting $f_{\rm gsm}$ (after five iterations) and $f_{\rm tsm}$.

are the same.

We found that even for very broad $g_{sm}$ functions the iterative scheme performed well. Figure 6 shows the density residuals in the construction of an E5 Kuzmin–Kutuzov model with $q = 2$ and $s^2_{max} = 0.9$. The computation was continued until $|\rho_i/\rho_m| < 10^{-3}$, which occurred after five iterative steps. Even for this 'fat' model the residuals decrease rapidly. The resulting $f_{gsm}$ is also shown, compared to the thin-orbit distribution function $f_{tsm}$.

### 3.5 Velocity moments

When a distribution function $f_{sm} = f_{gsm}c_g g_{sm}$ for a given density $\rho_m$ in a potential $V$ has been found, we are most often also interested in the observable kinematical properties of the resulting dynamical model. These follow by taking the appropriate velocity moments of $f_{sm}$, followed by a line-of-sight integration. It turns out that the intrinsic velocity moments can be found easily by a slight extension of our algorithm.

The intrinsic velocity moments are defined as

$$\rho_m \langle v_\lambda^i v_\phi^j v_\nu^k \rangle = \iiint v_\lambda^i v_\phi^j v_\nu^k f_{sm} \, dv_\lambda dv_\phi dv_\nu, \qquad (3.17)$$

The expressions for the velocity components at a point $(\lambda, \nu)$ along an orbit with turning points $(\nu_0, \lambda_1, \lambda_2)$ are given in equation (3.2) of ZH. Upon transformation to our variables $(s, t, u, x)$ we find

$$\begin{aligned} v_\lambda^2 &= 2(\lambda+\alpha)\frac{(1-x+xu)}{(1+t)^2}(s^2-t^2)U[\lambda, \lambda_1, \lambda_2, \nu_0], \\ v_\phi^2 &= 2(\lambda+\alpha)\frac{u(1-s^2)}{(1+t)^2}U[-\alpha, \lambda_1, \lambda_2, \nu_0], \qquad (3.18) \\ v_\nu^2 &= 2(\lambda-\nu)\frac{(1+xt)^2-(1-x)^2 s^2}{(1+t)^2}(1-u)U[\nu, \lambda_1, \lambda_2, \nu_0]. \end{aligned}$$

Hence, if we insert $v_\lambda^i v_\phi^j v_\nu^k$ in our equation (3.10), after use of the transformation (3.18), we find the contribution to the required moment. Since the $U$-functions occur in $U^*$, they have already been evaluated, so calculating the velocity moments in parallel with carrying out the iteration to get $f_{sm}$ adds very little to the required CPU time.

As an example, we have constructed two E5 Kuzmin–Kutuzov models with $q = 0$ and $s^2_{max} = 0.5$ and 0.7, respectively. In five iterations the residual density is less than $10^{-3}\rho_m$. The velocity moments are shown in Figure 7 (cf. Figures 5 and 7 in ZH). As expected, the dispersion in the $\nu$- and $\phi$-directions decrease while the 'radial' dispersion increases. The $s^2_{max} = 0.7$ model has the largest ratio of $\langle v_\lambda^2 \rangle / \langle v_\phi^2 \rangle$, which is of the order of 0.25.

### 3.6 Numerical implementation

We have written a code to implement the iterative scheme defined in equations (2.22) and (2.23). The density residuals and the terms in the series for $f_{gsm}$ are calculated on a $(\lambda, \nu)$ grid that doubles as a $(\lambda_m, \nu_0)$ grid. The quantities are interpolated using splines as their values are also needed in between mesh points to evaluate the $s$- and $t$-integrals. The actual integrations are carried out using Romberg integration after switching to more appropriate integration variables instead of $s$ and $u$ to remove the square roots from the denominators.

The $(\lambda, \nu)$ grid points have to be chosen with some care. We need to cover the full $\lambda$ domain in order to prevent boundary errors in the determination of the residual density. This is accomplished by using a grid that is linear in the variable $t_\lambda$ defined by

$$\lambda(t_\lambda) + \alpha = (\lambda_H + \alpha)\frac{(2t_\lambda)^p}{(2-2t_\lambda)^{p'}}, \qquad 0 < t_\lambda < 1, \qquad (3.19)$$

where $\lambda_H$ is the value of $\lambda$ in the middle of the grid, $p$ determines the resolution near $\lambda = -\alpha$ and $p'$ is chosen to match the large-radii behaviour of the distribution function. Most of the computations in this paper were done using $p = 3$, $p' = 0.25$ and $\lambda_H = 4$.

When the model is very flattened towards the equatorial plane, its associated prolate spheroidal coordinate system is very elongated along the $z$-axis. The net effect of this opposite elongation is that the density and distribution function are sharply peaked near $\nu = -\gamma$. We therefore use a $\nu$-grid that is linear in the variable $u_\nu$, where

$$\nu(u_\nu) + \gamma = A\left[\frac{(1+u_\nu)^p}{B + (1+u_\nu)^{p-1}} - \frac{1}{B+1}\right]. \qquad (3.20)$$

for some $p \geq 1$; $B$ is set to $(\frac{7}{4})^{p-1}$. The parameter $A$ is determined so that $\nu(1) = -\alpha$. We have found this substitution to be adequate for models as flattened as E7.

The thin-orbit distribution function is computed from the density residuals using (2.18). A straightforward implementation of (3.10) is feasible, but a single iteration step on a $(\lambda, \nu)$ grid of 50x50 would take about 100 minutes CPU time on an HP735. There is a faster way to implement the algorithm. The only part of the integral in (3.10) that changes in successive iterations is the $f_{tsm}\{\rho_i\}$ factor, which does not depend on $s$. We therefore exchange the order of integration, and carry out the $s$-integral first. It is

$$T_g(t, u, x) = \int_{t^2}^{s^2_{max}} ds^2 \, \frac{w_2(s, t, u) g_{sm}(s)}{\sqrt{s^2 - t^2}}, \qquad (3.21)$$

and can be tabulated before the first iteration. Using a 81x27x100 grid for $(t, u, x)$, the initialization stage takes 5 minutes CPU time, or 10 minutes if the velocity moments have to be computed as well. Each iteration step then simplifies to

$$\rho_{i+1} = \rho_i - \int_0^1 du \, w_1(u) \int_{-s_{max}}^{s_{max}} dt \, T_g(t, u, x) f_{tsm}\{\rho_i\}, \qquad (3.22)$$

which takes about 1 CPU minute to complete. The program is written in Fortran. The number of iteration steps to reach a relative accuracy better than $10^{-3}$ is 1 for $s_{max} = 0.1$, 3 for $s_{max} = 0.5$ and 5 for $s_{max} = 0.7$ in the case of $q = 0$ models. It is smaller when $q > 0$.

There is a further reduction possible in the limiting cases of a spherical or a disk galaxy. These are described in Sections 4.4 and 4.5, respectively.

## 4 SPECIAL CASES

Approximate solutions of the fundamental integral equation (3.10) can be found by analytic means at large radii, and near the foci of the spheroidal coordinates. We consider them in turn, and then show how the scheme simplifies for models with sharply peaked $g_{\rm sm}$ functions, and in the spherical and disk limits.

**Figure 7.** The intrinsic velocity moments for the E5 Kuzmin–Kutuzov thin-orbit model (solid) and $q = 0$ models with $s_{\max} = 0.5$ (dotted) and 0.7 (dashed lines).

### 4.1 Large radii

The relation (3.10) for the density residuals simplifies considerably in the limit of large radii, i.e., $\lambda \to \infty$ so that $x \to 0$. The functions $w_1$ and $w_2$ then reduce to

$$w_1(u) = \frac{4\sqrt{2}}{\sqrt{u(1-u)}},$$
$$w_2(s,t,u) = \frac{U^*(s,t,u)c_g(t,u)}{(1+t)^{3/2}}. \quad (4.1)$$

In order to evaluate the various third-order divided differences that appear in the definitions (2.15) and (3.4) for $U^*$ and $c_g$, respectively, we assume that $s_{\max} < 1$, so that the orbits that contribute to the density at $(\lambda, \nu)$ have $\lambda_2 \geq \lambda \geq \lambda_1 = \lambda_m(1-s) \geq \lambda_m(1-s_{\max}) \gg -\alpha$. The details of the calculations are given in Appendix B. We find that in this case neither $U^*$ nor $c_g$ depends on $u$ at large $\lambda$, and that

$$U^* c_g \simeq \frac{GM}{2^{5/2} C_g} \frac{(1+t)}{\lambda} L(s,t), \quad (4.2)$$

where $L(s,t)$ is the elementary function given in equation (B4), and the constant $C_g \geq 1$ is defined in equation (B10). $C_g = 1$ in the thin orbit limit $g_{\rm sm} = \delta(s^2)$.

We consider a flattened density $\rho_i$ that falls off as a power of $\lambda$ at large radii:

$$\rho_i(\lambda, \nu) \simeq \frac{\rho_i^\nu(\nu)}{\lambda^p}, \quad (4.3)$$

for some (positive) value of $p$. Selfconsistent models with finite total mass must have $p > 3/2$. By Kuzmin's formula, such models must also have $p \leq 2$ (e.g., de Zeeuw, Peletier & Franx 1986). Non-consistent densities $\rho_m$ may fall off steeper than this. Substitution of the form (4.3) in expression (2.18) for $f_{\rm tsm}\{\rho_i\}$ and use of the approximations (B1) then shows that $f_{\rm tsm} \propto \lambda_m^{1-p}$ times a function of $\nu_0$ (cf ZH,

eq. [2.56]). Transformation to the variables $(s,t,u)$ gives

$$f_{\rm tsm}\{\rho_i\} \simeq \frac{(1+t)^{p-1}}{\lambda^{p-1}} \frac{f_{\rm tsm}^\nu(u)}{\pi^2 GM}, \quad (4.4)$$

where

$$f_{\rm tsm}^\nu = \left[ \rho_i^\nu(-\alpha) + \sqrt{u} \int_0^u \frac{[d\rho_i^\nu(u')/du']\,du'}{\sqrt{u-u'}} \right]. \quad (4.5)$$

Thus, at large radii the thin orbit distribution function becomes a product of a function of $t$ and a function of $u$.

Substitution of all the above approximations in the basic relation (3.10) shows that for $\lambda \gg -\alpha$ the triple integration over $s, t$ and $u$ reduces to the product of an integral over $u$ times a double integral over $s$ and $t$:

$$\rho_{i+1} \simeq \rho_i - \frac{L_g(p)}{\pi C_g} \frac{1}{\lambda^p} \int_0^1 \frac{du\, f_{\rm tsm}^\nu(u)}{\sqrt{u(1-u)}}, \quad (4.6)$$

where we have defined

$$L_g(p) = \frac{1}{\pi} \int_0^{s_{\max}^2} ds^2 g_{\rm sm}(s) \int_{-s}^s \frac{dt\, L(s,t)}{(1+t)^{\frac{3}{2}-p}\sqrt{s^2-t^2}}. \quad (4.7)$$

The $u$-integration in equation (4.6) can be carried out, and equals $\pi \rho_i^\nu(\nu)$. Since $L_g(p)$ is a constant, it follows that the triple integration over $f_{\rm tsm}\{\rho_i\}$ is proportional to $\rho_i^\nu(\nu)/\lambda^p$, i.e., it is proportional to $\rho_i$ itself:

$$\rho_{i+1} \simeq \rho_i \left[ 1 - \frac{L_g(p)}{C_g} \right]. \quad (4.8)$$

Thus, the residual density at large radii becomes smaller by a constant factor $[1 - L_g(p)/C_g]$ at each iteration step.

In the thin orbit limit $g_{\rm sm} = \delta(s^2)$, and we have $C_g = L_g(p) = 1$, so that $\rho_{i+1} = 0$ for all $i$. This is as it should be, since $f_{\rm gsm}$ equals $f_{\rm tsm}\{\rho_m\}$ exactly in this case. Equation (4.8) implies that for broadened functions $g_{\rm sm}$ the distribution function $f_{\rm gsm}$ at large values of $\lambda_m$ is given by

$$f_{\rm gsm}(\nu_0, \lambda_m) \simeq \frac{C_g}{L_g(p)} f_{\rm tsm}\{\rho_m\}, \quad (\lambda_m \gg -\alpha), \quad (4.9)$$

so that we can find it without iteration.

**Figure 8.** Limiting behaviour at large radii of various properties of models with a density fall-off $\rho_m \propto 1/\lambda^2$, and functions $g_{sm}$ with $q = 0$ (solid), 1 (dotted), and 2 (dashed), and values of $s_{max}$ between 0 and 1. Here $\lambda \sim r^2$, and $f_{gsm}$ becomes a constant factor $C_g/L_g(2)$ times the thin-orbit distribution function $f_{tsm}$ (see eq. [4.9]). Shown are, as a function of $s_{max}$: a) the factor $C_g/L_g(2)$, b) the intrinsic mean streaming motion $\langle v_\phi \rangle$ in units of the maximum possible streaming $\langle v_\phi \rangle_{tsm}$, assuming all stars have the same sense of rotation around the symmetry axis, c) the second moment $\langle v_\lambda^2 \rangle$ of the 'radial' velocity, in units of $GM/\sqrt{\lambda}$, and d) the second moments $\langle v_\phi^2 \rangle$ and $\langle v_\nu^2 \rangle$ in units of their values in the thin-orbit model. These results are valid for $\lambda_m(1 - s_{max}) \gg -\alpha$.

The values of the constants $C_g$ and $L_g(p)$ can be found by numerical evaluation of the integrals (B10) and (4.7). For sharply peaked $g_{sm}$ they can be approximated by expanding the integrands in powers of $s^2$, and evaluating term by term. This gives

$$C_g \simeq 1 + \langle s^2 g_{sm} \rangle + \tfrac{501}{256}\langle s^4 g_{sm} \rangle + \mathcal{O}(\langle s^6 g_{sm} \rangle),$$
$$L_g(p) \simeq 1 + (\tfrac{25}{16} - \tfrac{7}{8}p + \tfrac{1}{4}p^2)\langle s^2 g_{sm} \rangle$$
$$+ (\tfrac{1005}{128} - \tfrac{471}{128}p + \tfrac{227}{128}p^2 - \tfrac{11}{32}p^3 + \tfrac{1}{32}p^4)\langle s^4 g_{sm} \rangle$$
$$+ \mathcal{O}(\langle s^6 g_{sm} \rangle), \quad (4.10)$$

so that

$$\rho_{i+1} \simeq \rho_i \left[ (\tfrac{9}{16} - \tfrac{7}{8}p + \tfrac{1}{4}p^2)\langle s^2 g_{sm} \rangle \right.$$
$$\left. + (\tfrac{45}{32} - \tfrac{359}{128}p + \tfrac{195}{128}p^2 - \tfrac{11}{32}p^3 + \tfrac{1}{32}p^4)\langle s^4 g_{sm} \rangle + \ldots \right]. \quad (4.11)$$

This shows again that for sharply peaked $g_{sm}$ its first moment is mostly responsible for the convergence of the iterative scheme.

In a similar fashion it is possible to derive approximations for the intrinsic velocity moments. Substituting approximations (B1) in (3.18), we find that the velocities can be written as a

$$v_\tau^2 = \frac{GM}{\sqrt{\lambda}} \ell_\tau(u) L_v^\tau(s,t), \qquad (\tau = \lambda, \nu, \phi), \quad (4.12)$$

where $L_v^\tau$ is given in (B12), and $\ell_\tau = 1, (1-u), u$ for $\tau = \lambda, \nu, \phi$, respectively. Substitution of (4.12) and (4.9) in (3.17) yields

$$\rho_m \langle v_\tau^n \rangle = \frac{L_{gv}^{\tau n}}{L_g(p)} \frac{1}{\pi \lambda^p} \left( \frac{GM}{\sqrt{\lambda}} \right)^{n/2} \int_0^1 \frac{du\, f_{tsm}^\nu(u) \ell_\tau(u)}{\sqrt{u(1-u)}}, \quad (4.13)$$

where we have introduced

$$L_{gv}^{\tau n}(p) = \frac{1}{\pi} \int_0^{s_{max}^2} ds^2 g_{sm}(s) \int_{-s}^{s} \frac{dt\, L_v^\tau(s,t)^{\tfrac{n}{2}} L(s,t)}{(1+t)^{\tfrac{3}{2}-p}\sqrt{s^2-t^2}}. \quad (4.14)$$

Again, the $u$-integration can be carried out. Since $L_v^\tau(0,0) = 1$ for $\tau = \nu, \phi$, the $u$-integral equals $\rho_m \langle v_\nu^n \rangle_{tsm}$ and $\rho_m \langle v_\phi^n \rangle_{tsm}$, respectively, which are the velocity moments of the thin-orbit model. For $\tau = \lambda$ the $u$-integral equals $\rho_m$, hence

$$\langle v_\tau^n \rangle = \langle v_\tau^n \rangle_{\mathrm{tsm}} \frac{L_{gv}^{\tau n}(p)}{L_g(p)}, \qquad (\tau = \nu, \phi),$$
$$\langle v_\lambda^n \rangle = \left(\frac{GM}{\sqrt{\lambda}}\right)^{n/2} \frac{L_{gv}^{\lambda n}(p)}{L_g(p)}.$$
(4.15)

For sharply peaked $g_{\mathrm{sm}}$ functions the dispersions and rotation velocity can therefore be approximated by

$$\langle v_\phi \rangle \simeq \langle v_\phi \rangle_{\mathrm{tsm}} \Big[ 1 + (\tfrac{3}{64} - \tfrac{1}{8}p)\langle s^2 g_{\mathrm{sm}} \rangle$$
$$+ (\tfrac{993}{8192} - \tfrac{175}{512}p + \tfrac{7}{128}p^2)\langle s^4 g_{\mathrm{sm}} \rangle + \ldots \Big];$$
$$\langle v_\lambda^2 \rangle \simeq \frac{GM}{\sqrt{\lambda}} \Big[ \tfrac{1}{8}\langle s^2 g_{\mathrm{sm}} \rangle + (\tfrac{67}{128} - \tfrac{1}{8}p)\langle s^4 g_{\mathrm{sm}} \rangle + \ldots \Big];$$
$$\langle v_\nu^2 \rangle \simeq \langle v_\nu^2 \rangle_{\mathrm{tsm}} \Big[ 1 + (-\tfrac{9}{8} + \tfrac{3}{4}p)\langle s^2 g_{\mathrm{sm}} \rangle$$
$$+ (-\tfrac{531}{128} - \tfrac{215}{64}p - \tfrac{3}{8}p^2)\langle s^4 g_{\mathrm{sm}} \rangle + \ldots \Big];$$
$$\langle v_\phi^2 \rangle \simeq \langle v_\phi^2 \rangle_{\mathrm{tsm}} \Big[ 1 + (\tfrac{1}{8} - \tfrac{1}{4}p)\langle s^2 g_{\mathrm{sm}} \rangle$$
$$+ (\tfrac{41}{128} - \tfrac{45}{64}p + \tfrac{1}{8}p^2)\langle s^4 g_{\mathrm{sm}} \rangle + \ldots \Big].$$
(4.16)

Figure 8 illustrates the behaviour of $C_g/L_g(p)$ for our choice of $g_{\mathrm{sm}}$, for $p = 2$. It also shows intrinsic velocity dispersions and rotational velocity in terms of the thin-orbit values. The above results show that for a given density distribution $\rho_{\mathrm{m}}(\lambda, \nu)$, and with $g_{\mathrm{sm}}$ a function of $s$ only, the kinematic properties at large radii do depend on $\lambda$ and $\nu$, but they follow from those in the thin orbit model by multiplication with a factor wich depends on $g_{\mathrm{sm}}$ and $p$ only.

Equation (4.8) shows that our iterative scheme will converge for any $g_{\mathrm{sm}}$ for which $0 \leq L_g(p)/C_g < 2$. A useful upper limit for $L_g(p)/C_g$ can be obtained by evaluating

$$\max_{0 \leq s \leq s_{\max}} \frac{1}{\pi h(s)} \int_{-s}^{s} \frac{\mathrm{d}t\, L(s,t)}{(1+t)^{\frac{3}{2}-p}\sqrt{s^2 - t^2}}. \qquad (4.17)$$

It is easily verified by numerical integration of (4.17) that $L_g/C_g < 2$ as long as $p < 9/2$. This includes all physically relevant cases, so our iterative scheme will always converge at large radii for all $g_{\mathrm{sm}}$ functions.

### 4.2  Behaviour near the foci

We now investigate the behaviour of the iterative scheme near the foci of the spheroidal coordinates, where $\lambda = \nu = -\alpha$. The variables $s, t, u$ and $x$ that appear in the basic relation (3.10) for the density residuals take their full range of values near the foci, but the triple integration over $s, t$ and $u$ can nevertheless be simplified, because the factors $f_{\mathrm{tsm}}\{\rho_i\}(t,u)$, $U^*(s,t,u)$, and $c_g(t,u)$ which appear in the integrand, all simplify.

The only orbits that contribute density at the foci in the thin-orbit model are those with $\lambda_m = \nu_0 = -\alpha$. Orbits with $\lambda_m > -\alpha$ lie on spheroidal shells which intersect the $z$-axis above the foci, while orbits with $\lambda_m = -\alpha$ and $\nu_0 < -\alpha$ are $z$-axis oscillations that do not reach the foci. It follows that $\rho_m(-\alpha, -\alpha)$ is determined exclusively by $f_{\mathrm{tsm}}\{\rho_m\}(-\alpha, -\alpha)$. The relation is (see ZH, eq. [2.44])

$$f_{\mathrm{tsm}}\{\rho_m\}(-\alpha, -\alpha) = \frac{\rho_m(-\alpha, -\alpha)\,[1+x_0]}{8\pi^2\sqrt{\gamma - \alpha}\,U[-\alpha, -\alpha, -\alpha, -\alpha]}, \quad (4.18)$$

where

$$x_0 = \frac{-\alpha - \nu_0}{\lambda_m - \nu_0} = \frac{xu(1+t)}{1 - x + xu(1+t)}. \qquad (4.19)$$

and $U[-\alpha, -\alpha, -\alpha, -\alpha] = U'''(-\alpha) > 0$. In the limit where $\lambda_m = \nu_0 = -\alpha$, the value of $x_0$ can still lie anywhere between 0 and 1, so that the factor in square brackets in equation (4.18) can take any value between 1 and 2, depending on the direction along which the *focal corner* in the $(\lambda_m, \nu_0)$-plane is approached. ZH refer to this property of $f_{\mathrm{tsm}}$ by saying that it has *radial behaviour* in the focal corner (Figure 3, and ZH Figure 2). Without this behaviour of $f_{\mathrm{tsm}}$ it would not reproduce the correct density $\rho_m(-\alpha, -\alpha)$.

Since the thin-orbit distribution function $f_{\mathrm{tsm}}\{\rho_i\}$ is proportional to $\rho_m(-\alpha, -\alpha)$, the residual density $\rho_1$ depends only on the local behaviour of the distribution function near the focal corner. This means that we can approximate $U^*$ in equation (3.10) by the constant value $[U'''(-\alpha)]^{3/2}$, so that it can be taken outside the triple integration.

We show in Appendix C that near the focal corner the function $c_g$ can be approximated as

$$c_g(\nu_0, \lambda_m) \simeq \frac{\sqrt{2(\gamma - \alpha)}}{\sqrt{U'''(-\alpha)}} \frac{1}{J_g(x_0)}, \qquad (4.20)$$

where $J_g(x_0)$ is a weighted integral of the function $g_{\mathrm{sm}}$, defined in equation (C7). As a result, $c_g$ also has radial behaviour near the focal corner, except in the thin-orbit limit when $J_g(x_0) \equiv 1$.

Upon substitution of these approximations in relation (3.10), we find

$$\rho_1(-\alpha, -\alpha) \simeq \rho_m(-\alpha, -\alpha)[1 - F_g(x)], \qquad (4.21)$$

where

$$F_g(x) = \frac{1}{\pi^2} \int_0^1 \frac{\mathrm{d}u}{\sqrt{u(1-u)(1-x+xu)}}$$
$$\times \int_0^{s^2_{\max}} \frac{\mathrm{d}s^2}{\sqrt{1-s^2}} \int_{-s}^{s} \frac{\mathrm{d}t}{(1+t)^{3/2}} \frac{g_{\mathrm{sm}}(s)}{\sqrt{s^2 - t^2}} \frac{(1+x_0)}{J_g(x_0)} \qquad (4.22)$$
$$\times \frac{\big[[1-x+xu(1+t)]^2 - (1-x)^2 s^2\big]}{\sqrt{(1+tx)^2 - (1-x)^2 s^2}\sqrt{1-x+xu(1+t)}},$$

and we still have to substitute relation (4.19) for $x_0 = x_0(x, u, t)$. When $g_{\mathrm{sm}}(s) = \delta(s^2)$ the integration over $t$ and $s$ gives $\pi$ since then $J_g(x_0) \equiv 1$, and so does the remaining integral over $u$, so that then $F_g(x) \equiv 1$ and the residual $\rho_1 \equiv 0$, as it should be for the thin-orbit model. However, when $g_{\mathrm{sm}}$ is not infinitely sharply peaked, the value of $\rho_1$ at the foci ($\lambda = \nu = -\alpha$) depends on the direction of approach, i.e., on the value of $x$. Thus, the residual density has *radial behaviour* near the foci.

The triple integral (4.22) requires numerical evaluation for $0 < x < 1$ and general $g_{\mathrm{sm}}$. We have found it to be a slowly varying monotonic function of $x$ for our choice (3.6) of functions $g_{\mathrm{sm}}(s)$ (see Figure 9). It is bounded by the values $F_g(0)$ and $F_g(1)$. We show in Appendix D that the triple integration reduces to a single integral for $x = 0$ and $x = 1$, and that the remaining integrals can be found explicitly for our set of functions (3.6). By combining equations (C9) and

**Figure 9.** The behaviour of $(1 + x_0)/F_g(x_0)$ for (a) $q = 2$, (b) $q = 1$ and (c) $q = 0$, as calculated with our iterative scheme, for $s_{\max}^2 = 0, \ldots, 0.9$ in steps of 0.1. The factor $1/F_g(x_0)$ decreases monotonically with $s_{\max}$. At most 10 iterations were computed for each model, regardless of the achieved accuracy. The scheme did not converge for the $(q = 0, s_{\max}^2 = 0.9)$ model. In (d) and (e) the results (crosses) from (a)–(c) are compared to the analytical value for $q = 0$ (solid), 1 (dotted) and 2 (dashed) as a function of $s_{\max}$. In (f) the lines $F_g(1) = 2$ (dotted) and $F_g(0) = 2$ (solid), which are in effect a relation between $q$ and $s_{\max}$, are plotted for our $g_{\text{sm}}$ functions. In the shaded area both $F_g(0) < 2$ and $F_g(1) < 2$, which is an indicator for convergence of the iterative scheme.

(D6) we find, for $q > 0$:

$$F_g(0) = \frac{{}_2F_1(\frac{3}{4}, \frac{5}{4}; 2+q; s_{\max}^2)}{{}_2F_1(\frac{1}{4}, \frac{3}{4}; 2+q; s_{\max}^2)},$$
$$F_g(1) = \frac{{}_2F_1(1, 1; 2+q; s_{\max}^2)}{{}_2F_1(\frac{1}{2}, 1; 2+q; s_{\max}^2)},$$
(4.23)

while for $q = 0$ we have (cf. eqs [C10] and [D8])

$$F_g(0) = \frac{3\left[K(k) - (1+s_{\max})E(k)\right]}{(1+s_{\max})\left[E(k) - (1-s_{\max})K(k)\right]},$$
$$F_g(1) = -\frac{(1+\sqrt{1-s_{\max}^2})}{2s_{\max}^2} \ln(1-s_{\max}^2),$$
(4.24)

where $K$ and $E$ are the complete elliptic integrals of the first and second kinds, and $k^2 = 2s_{\max}/(1+s_{\max})$. For each $q \geq 0$ these functions increase monotonically with increasing $s_{\max}$. In the limit $s_{\max} \to 1$ each function reaches a finite value,

$$F_g(0) = \frac{q + \frac{3}{4}}{q}, \qquad F_g(1) = \frac{q + \frac{1}{2}}{q}, \qquad (4.25)$$

provided $q > 0$. However, $F_g(0)$ and $F_g(1)$ diverge logarithmically when $q = 0$ and $s_{\max} \to 1$, although their ratio approaches $3/2$, in agreement with the result $F_g(0)/F_g(1) \to (q+\frac{3}{4})/(q+\frac{1}{2})$ which follows from equation (4.25). For given $q$ and $s_{\max}$ the total relative variation of $F_g(x)$ between $x = 0$ and $x = 1$ is therefore never larger than 1.5. Figure 9 illustrates the behaviour of $F_g(0)$ and $F_g(1)$ as a function of $s_{\max}$ for $q = 0, 1$ and 2.

The radial behaviour of the residual density $\rho_1$ at the foci of the spheroidal coordinates, and the fact that the distribution function $f_{\rm sm} = f_{\rm gsm}\tilde{g}_{\rm sm}$ is determined by the local density distribution when $s_{\max} < 1$, means that $f_{\rm gsm}(\nu_0, \lambda_m)$ must have radial behaviour near the focal corner $\nu_0 = \lambda_m = -\alpha$, so that it must be of the form

$$f_{\rm gsm}\{\rho_m\}(-\alpha,-\alpha) = f_{\rm tsm}\{\rho_m\}(-\alpha,-\alpha)K_g(x_0)$$
$$= \frac{\rho_m(-\alpha,-\alpha)(1+x_0)K_g(x_0)}{8\pi^2\sqrt{\gamma-\alpha}\,U'''(-\alpha)}, \qquad (4.26)$$

with $K_g$ a function of the variable $x_0$ defined in equation (4.19). Since $x = 0$ corresponds to $x_0 = 0$, and since similarly $x = 1$ corresponds to $x_0 = 1$, it follows that

$$K_g(0) = \frac{1}{F_g(0)}, \qquad K_g(1) = \frac{1}{F_g(1)}. \qquad (4.27)$$

This suggests — but does not prove — that $K_g(x_0)$ has a modest variation with $x_0$. It can be computed as follows, at least in principle. We insert $K_g(x_0)$ as a factor in the triple integral on the right-hand side of equation (4.22), and put the left-hand side equal to 1 for $0 \leq x \leq 1$. Transformation of the variables $(u, t)$ to $(x_0, t)$ then allows one to carry out the $t$ and $s$-integrals. This leaves a one-dimensional integral equation for the function $K_g(x_0)$. In practice this must be solved numerically, and it is in fact easier to simply use our iterative scheme. We have applied it to compute $K_g(x_0)$ for our functions $g_{\rm sm}$ with $q = 0, 1$ and $s_{\max} = 0, 0.3, 0.5, 0.7$ and 0.9. The resulting functions are shown in Figure 9a,b,c. They are very nearly linear up to values of $s_{\max}$ as large as 0.7. This means that to good approximation we can take

$$K_g(x_0) \simeq \frac{1-x_0}{F_g(0)} + \frac{x_0}{F_g(1)}, \qquad (4.28)$$

so that we can find the local behaviour of the distribution function near the focal corner without iteration.

The results presented in Figure 9d,e also show that the jump in the value of $f_{\rm gsm}$ at the focal corner is a function of $s_{\max}$. The magnitude of the jump going from $x_0 = 0$ to $x_0 = 1$ is $2K_g(1)/K_g(0) = 2F_g(0)/F_g(1)$, and hence varies from 2 when $s_{\max} = 0$ in the thin-orbit model to $(2q + \frac{3}{2})/(q + \frac{1}{2})$ when $s_{\max} \to 1$.

The convergence of the iterative scheme near the foci of the model cannot be derived by investigation of only the residual density $\rho_1$, as we have done in the above. However, equation (4.21) suggests that in cases where $0 \leq F_g(x) \leq 2$ also the higher order residuals $|\rho_i|$ with $i > 1$ will decrease in size, so that the scheme very likely will converge. This condition on $F_g(x)$ is met for all our functions $g_{\rm sm}$ when $q \geq 3/4$, and is also met for a large range of $s_{\max}$ when $-1 < q < 3/4$ (see Figure 9f).

The above results also hold when $s_{\max} = 1$ as long as $q > 0$. In this case there are orbits with arbitrarily large outer turning point $\lambda_2$ but low angular momentum that still provide density at the foci, but their contribution is vanishingly small. However, when $q = 0$ and $s_{\max} = 1$ this is no longer so, and our derivation of the approximate relation (4.21) is invalid. The density close to the foci depends on the details of the distribution of non-local orbits.

We remark that all self-consistent oblate models with density $\rho_m$ in a potential $V$ of the form (2.2) have distribution functions with radial behaviour in the focal corner, unless $q \leq 0$ and $s_{\max} = 0$. From our analysis it is not clear whether the radial behaviour is still present in the latter case. However, the two-integral distribution function $f_{\rm sm}(E, L_z)$ is generally well-behaved at the focal corner (Figure 4). When written in the form (2.16), it leads to a function $g_{\rm sm}$ which does not depend on $s$ alone. At the focal corner it is identical to our $q = 0$, $s_{\max} = 1$ function, and along $\nu_0 = -\alpha$ it has $q < 0$ behaviour. Our iterative scheme with $f_{\rm tsm}$ as initial guess for $f_{\rm gsm}$ fails to converge in this case.

### 4.3  Models with peaked $g_{\rm sm}$ functions

We have seen that the convergence of the iterative scheme is determined by the moments of the $g_{\rm sm}$ function. For models with sharply peaked $g_{\rm sm}$ functions only the first moment is significant; higher moments can be neglected. In this case only a single iteration is required to approximate $f_{\rm gsm}$ to high accuracy.

The expression for the residual density $\rho_1$ simplifies considerably. Since $|t| \leq s \leq s_{\max}$ and $s_{\max}$ is small, the $w_2(s, t, u)$ function in (3.10) can be expanded in a Taylor series in $t$ and $s^2$:

$$w_2(s,t,u) \simeq w_2 + \frac{\partial w_2}{\partial s^2}s^2 + \frac{\partial w_2}{\partial t}t + \frac{1}{2}\frac{\partial^2 w_2}{\partial t^2}t^2 + \ldots, \quad (4.29)$$

where the derivatives are evaluated at $s = t = 0$. Similarly, $f_{\rm tsm}(t, u)$ can be written as a Taylor series in $t$. The $s$- and $t$-integrations can be carried out. The $w_2(0, 0, u)f_{\rm tsm}(0, 0)$ term yields the model density $\rho_0$, hence the residual density $\rho_1$ is

$$\rho_1 = \langle s^2 g_{\rm sm}\rangle \delta\rho \qquad (4.30)$$

**Figure 10.** The function $\delta f/f_{\rm tsm}$ for an E5 Kuzmin–Kutuzov model as obtained from a model with $s_{\rm max} = 0.1$ and (a) $q = 0$, (b) $q = 1$ and (c),(d) $q = 2$ (shaded surfaces). The difference of $\delta f/f_{\rm tsm}$ obtained from (a)–(c) $s_{\rm max} = \sqrt{0.1}$ and $s_{\rm max} = 0.1$ and (d) $q = 0$ and $q = 2$ models are shown as wireframe surfaces.

with

$$\delta\rho = -\pi \int_0^1 du\, w_1(u) \left[ \left( \frac{\partial w_2}{\partial s^2} + \frac{\partial^2 w_2}{4\partial t^2} \right) f_{\rm tsm} + w_2 \frac{\partial^2 f_{\rm tsm}}{4\partial t^2} \right], \quad (4.31)$$

with $w_2$ and $f_{\rm tsm}$ evaluated at $s = t = 0$. Hence $\delta\rho$ is independent of the shape of $g_{\rm sm}$. This relation even holds if $g_{\rm sm}$ also depends on $\lambda_m$ and $\nu_0$. The resulting $f_{\rm gsm}$ can be approximated by

$$f_{\rm gsm} \simeq f_{\rm tsm} + \langle s^2 g_{\rm sm} \rangle \delta f \quad (4.32)$$

where $\delta f = f_{\rm tsm}\{\delta\rho\}$ is also independent of $g_{\rm sm}$. Thus, all models with sharply peaked $g_{\rm sm}$ functions can effectively be described by the one-parameter family (4.32).

We have computed $\delta f$ for an E5 Kuzmin–Kutuzov model with $q = 0, 1, 2$ and $s_{\rm max}^2 = 0.01$ and $0.1$. The results are displayed in Figure 10 as shaded surfaces. To verify that $\delta f$ is indeed independent of $g_{\rm sm}$, i.e., of $q$ and $s_{\rm max}$, we compare the difference of two $\delta f$-functions obtained for a different set of $(q, s_{\rm max})$ with $\delta f$ itself. It is clear from Figure 10 that the differences are indeed much smaller, so that (4.32) holds to high accuracy.

### 4.4 Spherical limit

When $\gamma \to \alpha$, the prolate spheroidal coordinates $(\lambda, \nu, \phi)$ reduce to spherical coordinates $(r, \theta, \phi)$ with $r^2 = \lambda + \alpha$ (e.g., dZ). The potential (2.2) now is spherical, and equals $V(r) = -G(\lambda)$. The thin-orbit model reduces to the spherical model built exclusively with circular orbits (ZH, §IIg). Our iterative scheme will produce spherical models with a pre-assigned distribution of relative orbital thickness. We summarize briefly how the algorithm simplifies in this case.

In the limit $\gamma = \alpha$, we must have $\nu = \nu_0 = -\alpha$, so that $x = 0$. This reduces the expressions for $w_1$ and $w_2$ that appear in the basic relation (3.10) to the forms already given in equation (4.1). Furthermore, the thin-orbit function $f_{\rm tsm}$ becomes a function of $\lambda_m$ only. It can be evaluated without the need for integration since the density at radius $r_m$, say, is provided only by the circular orbits with radius $r_m$. We write $f_{\rm tsm}$ as (cf. ZH, corrected for a typographical error of a factor of 3)

$$f_{\rm tsm}\{\rho_i\}(r_m) = \frac{\rho_i(r_m)}{\pi^2 r_m \kappa_0^2(r_m)}, \quad (4.33)$$

where $\kappa_0^2$ is the epicyclic frequency, given by $r\kappa_0^2(r) =$

$3V'(r) + rV''(r)$, and $r_m$ is defined by $r_m^2 = \lambda_m + \alpha$. Also,

$$c_g = \frac{r_m^3}{S_g(r_m)}, \tag{4.34}$$

where $S_g$ is the integral in the denominator of the definition (3.4). It contains the factor $|\partial J_\lambda/\partial s^2|$ which is given in equation (A3), and must be evaluated with $-\gamma = \nu_0 = -\alpha$. Finally, the function $U^*$ defined in equation (2.15) reduces to

$$U^* = \frac{U[-\alpha, \lambda_1, \lambda_1, \lambda_2] U[-\alpha, \lambda_1, \lambda_2, \lambda_2]}{\sqrt{U[-\alpha, \lambda_1, \lambda, \lambda_2]}}. \tag{4.35}$$

We write $r_1^2 = \lambda_1 + \alpha$ and $r_2^2 = \lambda_2 + \alpha$, so that $r_1$ and $r_2$ are the inner and outer radius reached by an orbit, and $r_m^2 = \frac{1}{2}(r_1^2 + r_2^2)$. Since $U(\lambda) = -(\lambda + \alpha)^2 G(\lambda) = r^4 V(r)$, it follows from the definition (2.9) of the divided differences that the function $U^*$ is a combination of $V(r)$ and its derivative $V'(r)$ evaluated at $r$, $r_1$ and $r_2$. These are related to the variables $s$ and $t$ by

$$s = \frac{r_2^2 - r_1^2}{2r_m^2}, \qquad t = \frac{r^2 - r_m^2}{r_m^2}. \tag{4.36}$$

so that $U^* = U^*(r, s, t)$, $c_g = c_g(r, t)$ and $f_{\mathrm{tsm}} = f_{\mathrm{tsm}}(r, t)$. This means that we can carry out the $u$-integration in equation (3.10) to give $\pi$.

Substitution of the above results and exchange of the order of the $s$- and $t$-integrations, then reduces the basic relation (3.10) for the residuals $\rho_i$ to

$$\rho_{i+1}(r) = \rho_i(r) - \frac{4\sqrt{2}}{\pi} \int_{-s_{\max}}^{s_{\max}} \mathrm{d}t \, \frac{H_g(r,t)}{(1+t)^{3/2}} \frac{r_m^2 \rho_i(r_m)}{S_g(r_m)\kappa_0^2(r_m)}, \tag{4.37}$$

where $r_m = r(1+t)^{-1/2}$, and we have defined

$$H_g(r, t) = \int_{t^2}^{s_{\max}^2} \frac{\mathrm{d}s^2}{\sqrt{s^2 - t^2}} g_{\mathrm{sm}}(s) U^*(r, s, t). \tag{4.38}$$

Hence, we can choose a function $g_{\mathrm{sm}}$, evaluate $H_g(r, t)$ once, and then compute $\rho_{i+1}$ at each radius $r$ as a single weighted integral over $\rho_i(r)$. In practice it is convenient to use $r_m$ rather than $t$ as the integration variable in equation (4.37). No further work is needed to find the entire distribution function $f_{\mathrm{sm}} = f_{\mathrm{gsm}} \tilde{g}_{\mathrm{sm}}$ of the model, because it is given by

$$f_{\mathrm{sm}}(r_m, s) = \frac{g_{\mathrm{sm}}(s)}{\pi^2 r_m S_g(r_m) \kappa_0^2(r_m)} \sum_{i=0}^{\infty} \rho_i(r_m). \tag{4.39}$$

Calculation of the distribution function is thus considerably faster than in the flattened case, where computation of each $\rho_i$ requires a double integration over a distribution function $f_{\mathrm{tsm}}$ which itself is evaluated as a quadrature.

We remark that the arguments $r_m$ and $s$ of the above spherical distribution function each depend on the classical integrals of motion $E$ and $L^2$. The relations follows from

$$\begin{aligned} E &= \frac{r_2^2 V(r_2) - r_1^2 V(r_1)}{r_2^2 - r_1^2}, \\ L^2 &= 2 r_1^2 r_2^2 \frac{V(r_2) - V(r_1)}{r_2^2 - r_1^2}. \end{aligned} \tag{4.40}$$

with $r_1^2 = r_m^2(1-s)$ and $r_2^2 = r_m^2(1+s)$. For given $r_m$ and $s$ these relations must generally be inverted numerically to give the associated $E$ and $L^2$.

### 4.5 Disks

The iterative scheme (2.22)–(2.23) can also be applied in the limit where the density flattens to a circular disk with surface density $\Sigma_m(\lambda) = \int \rho_m \, \mathrm{d}z$, where $\lambda + \alpha = R^2$. The only orbits that can now be populated are those in the equatorial plane $z = 0$, so that we must have $\nu = \nu_0 = -\gamma$ and hence $u = 1$ in our fundamental relation (3.10). This leads to a number of simplifications.

When $u \uparrow 1$, the thin-orbit function $f_{\mathrm{tsm}}$, defined in equation (2.19) can be approximated as

$$\begin{aligned} f_{\mathrm{tsm}}\{\rho_i\} &= \frac{1}{8\pi^2\sqrt{\lambda_m + \gamma}} \frac{1}{U[-\gamma, \lambda_m, \lambda_m, \lambda_m]} \\ &\times \frac{\sqrt{U[-\gamma, -\alpha, \lambda_m, \lambda_m]}}{\sqrt{U[-\gamma, -\gamma, \lambda_m, \lambda_m]}} \lim_{u \uparrow 1} \left\{ \frac{\partial}{\partial u} \int_0^u \frac{\rho_i(\lambda_m, u') \, \mathrm{d}u'}{\sqrt{u - u'}} \right\}, \end{aligned} \tag{4.41}$$

where $\rho_i(\lambda_m, u') = \Sigma_i(\lambda_m) \delta(u' - 1)$. The function $c_g$ now is

$$c_g = \frac{(\lambda_m + \alpha)\sqrt{\lambda_m + \gamma}}{D_g(\lambda_m)}, \tag{4.42}$$

with $D_g$ the integral in the denominator of equation (3.4), evaluated at $\nu = \nu_0 = -\gamma$. The function $U^*$ reduces slightly, and becomes

$$\begin{aligned} U^* &= U[-\gamma, \lambda_1, \lambda_1, \lambda_2] \, U[-\gamma, \lambda_1, \lambda_2, \lambda_2] \\ &\times \sqrt{\frac{U[-\gamma, -\gamma, \lambda_1, \lambda_2]}{U[-\gamma, \lambda_1, \lambda_2, \lambda_2] \, U[-\gamma, -\alpha, \lambda_1, \lambda_2]}}. \end{aligned} \tag{4.43}$$

If we write $R_1^2 = \lambda_1 + \alpha$, $R_2^2 = \lambda_2 + \alpha$, $R_m^2 = \lambda_m + \alpha$, and use $U(\lambda) = -(\lambda + \alpha)(\lambda + \gamma)G(\lambda) = R^2(R^2 + \gamma - \alpha)V(R)$, with $V(R)$ the potential in the equatorial plane, then we can express all the above third-order divided differences in terms of $V$ and its derivatives. In particular, we have

$$\begin{aligned} U[-\gamma, -\alpha, \lambda, \lambda] &= \tfrac{1}{2}\Omega_0^2(R), \\ U[-\gamma, \lambda, \lambda, \lambda] &= \tfrac{1}{8}\kappa_0^2(R), \end{aligned} \tag{4.44}$$

with $\kappa_0$ the epicyclic frequency defined in Section 4.3, and $\Omega_0$ the circular frequency, given by $R\Omega_0^2(R) = V'(R)$. The cylindrical radius $R$ coincides with the spherical radius $r$ in the equatorial plane, so equation (4.36) can be used to find $U^* = U^*(R, s, t)$, $c_g = c_g(R, t)$ and $f_{\mathrm{tsm}} = f_{\mathrm{tsm}}(R, t)$.

We substitute the above approximations in equation (3.10), and integrate it over $z$ in order to obtain the basic relation for the residuals in the surface density. The $u$-integration can be carried out, and we are left with

$$\Sigma_{i+1}(R) = \Sigma_i(R) - \frac{4}{\pi} \int_{-s_{\max}}^{s_{\max}} \frac{\mathrm{d}t \, \tilde{H}_g(R, t) R_m^2 \Omega_0(R_m) \Sigma_i(R_m)}{(1+t)^{3/2}(1+xt)^{1/2} D_g(R_m) \kappa_0^2(R_m)}, \tag{4.45}$$

where $R_m = R(1+t)^{-1/2}$, $x = (\gamma - \alpha)/(R_m^2 + \gamma - \alpha)$, and we have defined

$$\tilde{H}_g = \int_{t^2}^{s_{\max}^2} \mathrm{d}s^2 g_{\mathrm{sm}}(s) U^*(R, s, t) \frac{\sqrt{(1+xt)^2 - (1-x)^2 s^2}}{\sqrt{s^2 - t^2}\sqrt{1 - s^2}}. \tag{4.46}$$

Just as in the spherical limit, the iterative scheme (2.22)–(2.23) simplifies considerably. We can choose a function $g_{\mathrm{sm}}(s)$, integrate it to get $\tilde{H}_g(R, t)$, and then evaluate $\Sigma_{i+1}$ at radius $R$ by a single quadrature. We have written $g_{\mathrm{sm}}$ here as a function of $s$ alone, but a dependence on $R_m$ can be included easily.

The three-dimensional distribution function of an infinitesimally thin disk can be written as

$$f_{\rm sm}(J_\lambda, J_\phi, J_\nu) = f_{\rm disk}(J_\lambda, J_\phi)\frac{\delta(J_\nu)}{2\pi}, \qquad (4.47)$$

where the division by $2\pi$ ensures that $f_{\rm disk}(J_\lambda, J_\phi)$ is the proper distribution function for the disk considered as a two-dimensional system. It follows from equations (2.22) and (4.45) that our scheme gives $f_{\rm disk}$ as

$$f_{\rm disk}(R_m, s) = \frac{\Omega_0(R_m)}{\pi\kappa_0^2(R_m)}\frac{g_{\rm sm}(s)}{D_g(R_m)}\sum_{i=0}^{\infty}\Sigma_i(R_m). \qquad (4.48)$$

The distribution function (4.48) can be written as a function of $E$ and $J_\phi = L_z$ by use of equation (4.40), with $r$ replaced by $R$, and $L^2$ by $L_z^2$.

We conclude that our iterative scheme provides a swift way to construct distribution functions for spheres and disks with a chosen distribution of the relative weights of orbits with different 'thickness' $r_2^2 - r_1^2$ and mean radial extent $r_m^2$. Based on our results in Sections 4.1 and 4.2, we expect the scheme to converge quickly, except for choices of $g_{\rm sm}$ that put a lot of weight on radial orbits, i.e., that have $g_{\rm sm}(1) > 0$.

## 5 CONCLUDING REMARKS

We have presented a simple numerical scheme for the construction of three-integral distribution functions for self-consistent and non-consistent oblate galaxy models with a potential of Stäckel form. The intrinsic velocity moments can be computed simultaneously. The algorithm allows one to choose in advance the distribution of the inner and outer turning points of the short-axis tube orbits that are populated. It then derives the entire distribution function from the density distribution by means of an iterative process that starts from the explicitly known distribution function of the thin-orbit (maximum streaming) model, in which only the tubes with equal inner and outer turning points are occupied. We have shown that this scheme works well, and is capable of producing tangentially anisotropic models with a substantial radial velocity dispersion within a few iteration steps. The algorithm simplifies considerably in the spherical and disk limits.

Dehnen & Gerhard (1993) have shown that three-integral flattened models display a large variety of observable kinematic properties, which include the line-of-sight mean velocity and velocity dispersion, as well as the entire distribution of the line-of-sight velocity (the velocity profile), all as a function of projected position on the sky. The observable kinematics of the tangentially anisotropic models constructed here can be computed in a straightforward way by numerical integration of the velocity moments and the distribution function, all of which are given with high accuracy by the algorithm.

We have investigated three special cases where three-integral distribution functions can be found without iteration.

First, models that have modest radial dispersions can be approximated adequately by a one-parameter family of distribution functions, which is insensitive to the detailed shape of the assigned function $g_{\rm sm}$, but depends only on its first moment. We will use this family in a subsequent paper to investigate the stability of cold oblate models.

Second, the structure of the model near the foci of the prolate spheroidal coordinate system in which the equations of motion separate provides information on the convergence of the algorithm. When the function $g_{\rm sm}$ is chosen such that only a vanishingly small number of orbits with $L_z = 0$ and a large outer turning point are occupied, the density near the foci is determined locally, i.e., by stars on orbits that are very close to the $z$-axis oscillations that just reach the foci. We have derived the local behaviour of the distribution function in all such models, and we have shown by analysis of the first residual density that the algorithm is very likely to converge in these cases, as indeed found numerically. However, when $L_z = 0$ orbits with large outer turning points contribute significantly to the density at the foci — which occurs in the $f(E, L_z)$ model, and in strongly radially anisotropic models — our algorithm appears to have problems, at least when we take $f_{\rm tsm}$ as initial guess for $f_{\rm gsm}$. In view of Bishop's (1986) work, we expect that a similar iterative scheme can be used for such models, but with $f(E, L_z)$ as zeroth order distribution function.

Third, the distribution functions of the models also simplify at large radii. There they reduce to a known factor times the distribution function of the thin-orbit model, which can be calculated easily. The internal velocity moments similarly simplify at large radii. This is useful, as it allows a straightforward calculation of the observable kinematic properties in the outer regions of these anisotropic flattened models. We intend to do so in a future paper. Absorption line kinematic measurements of elliptical galaxies now reach beyond two effective radii, and a comparison of these data with anisotropic models of the kind produced by our algorithm should provide further constraints on the presence and shape of a massive dark halo and the dynamics of the outer luminous regions of these systems (e.g., Carollo et al. 1995).

Finally, we remark that the iterative scheme is not restricted to oblate galaxy models. Prolate Stäckel models have two families of tube orbits, and the thin-orbit solutions have been described by Hunter et al. (1990). By applying our algorithm separately to the two tube orbit families, we can construct models with thick tubes. Triaxial Stäckel models contain three families of tube orbits as well as box orbits. The thin-orbit distribution functions for all three tube families can be found by simple quadratures (Hunter & de Zeeuw 1992; Arnold, de Zeeuw & Hunter 1994), and these can again be thickened by our algorithm. The tube orbits account for part of the density; the remainder must be reproduced by the box orbits. Their distribution function can be found by (numerically) solving a set of linear equations after the tube orbits have been populated. This last construction step is the same in thin and thick orbit models. Work on these triaxial models is in progress.

It is a pleasure to thank Richard Arnold and Marijn Franx for useful discussions and for comments on the manuscript.

## APPENDIX A: THE FUNCTION $\partial J_\lambda / \partial s^2$

In order to evaluate $c_g$ defined in equation (3.4), we need to calculate $|\partial J_\lambda / \partial s^2|$ at fixed $\nu_0$ and $\lambda_m$. The action $J_\lambda$ as a function of the turning points $(\nu_0, \lambda_1, \lambda_2)$ is defined in equation (2.11) as a single quadrature. Upon transformation to $(\nu_0, \lambda_m, \epsilon)$ we have

$$J_\lambda = \frac{\sqrt{2}}{\pi} \int_{\lambda_m - \epsilon}^{\lambda_m + \epsilon} d\lambda \frac{\sqrt{(\lambda - \lambda_m + \epsilon)(\lambda_m - \lambda + \epsilon)}}{(\lambda + \alpha)\sqrt{\lambda + \gamma}} \times \sqrt{(\lambda - \nu_0)} U[\nu_0, \lambda_m - \epsilon, \lambda, \lambda_m + \epsilon]. \quad (A1)$$

Since $s = \epsilon / (\lambda_m + \alpha)$, we have, at fixed $\lambda_m$,

$$\frac{\partial J_\lambda}{\partial s^2} = \frac{(\lambda_m + \alpha)^2}{2\epsilon} \frac{\partial J_\lambda}{\partial \epsilon}. \quad (A2)$$

The integrand in equation (A1) vanishes at the lower and upper limits of integration, so we can simply carry out the differentiation with respect to $\epsilon$ inside the integral. This is straightforward upon repeated use of the definition (2.9) of divided differences. The result can be written compactly as:

$$\frac{\partial J_\lambda}{\partial s^2} = \frac{(\lambda_m + \alpha)^2}{\pi \sqrt{2}} \int_{\lambda_1}^{\lambda_2} \frac{d\lambda}{(\lambda + \alpha)} \sqrt{\frac{\lambda - \nu_0}{\lambda + \gamma}} \sqrt{\frac{U[\nu_0, \lambda_1, \lambda, \lambda_2]}{(\lambda_2 - \lambda)(\lambda - \lambda_1)}} \times \left\{ 1 + \frac{(\lambda_2 - \lambda)(\lambda - \lambda_1) U[\nu_0, \lambda_1, \lambda_1, \lambda, \lambda_2, \lambda_2]}{U[\nu_0, \lambda_1, \lambda, \lambda_2]} \right\}. \quad (A3)$$

This is a function of $\nu_0$, $\lambda_1$ and $\lambda_2$, and hence depends on $\nu_0$, $\lambda_m$ and $\epsilon$, or $s$. We remark that the expressions for $J_\lambda$ and $\partial J_\lambda / \partial s^2$ are invariant under the exchange $\lambda_1 \leftrightarrow \lambda_2$. This means that both these functions are even in $s$, and hence are functions of $s^2$.

We found it convenient to evaluate $J_\lambda(\nu_0, \lambda_m, s)$ and $\partial J_\lambda / \partial s^2 (\nu_0, \lambda_m, s)$ by transformation to the integration variable $w$, defined as

$$w = \frac{2\lambda - \lambda_1 - \lambda_2}{\lambda_1 + \lambda_2} = \frac{\lambda - \lambda_m}{\epsilon} = \frac{t}{s}. \quad (A4)$$

Then $d\lambda = s(\lambda_m + \alpha) dw$, and the integration limits are $w(\lambda_1) = -1$ and $w(\lambda_2) = 1$. As a result

$$J_\lambda = \frac{\sqrt{2}}{\pi} s^2 (\lambda_m + \alpha) \sqrt{\lambda - \nu_0} \int_{-1}^{1} dw \sqrt{1 - w^2}$$
$$\times \frac{\sqrt{1 + (1 - x_0) sw}}{1 + sw} \sqrt{\frac{U[\nu_0, \lambda_1, \lambda, \lambda_2]}{\lambda + \gamma}},$$

$$\frac{\partial J_\lambda}{\partial s^2} = \frac{(\lambda_m + \alpha) \sqrt{\lambda_m - \nu_0}}{\pi \sqrt{2}} \int_{-1}^{1} \frac{dw}{\sqrt{1 - w^2}} \quad (A5)$$
$$\times \frac{\sqrt{1 + (1 - x_0) sw}}{1 + sw} \sqrt{\frac{U[\nu_0, \lambda_1, \lambda, \lambda_2]}{\lambda + \gamma}}$$
$$\times \left\{ 1 + \frac{s^2 (\lambda_m + \alpha)^2 (1 - w^2) U[\nu_0, \lambda_1, \lambda_1, \lambda, \lambda_2, \lambda_2]}{U[\nu_0, \lambda_1, \lambda, \lambda_2]} \right\},$$

where we still have to substitute $\lambda = \lambda_m + sw(\lambda_m + \alpha)$, $\lambda_1 = \lambda_m - s(\lambda_m + \alpha)$, and $\lambda_2 = \lambda_m + s(\lambda_m + \alpha)$. The quantity $x_0$ is defined in equation (4.19):

$$x_0 = \frac{-\alpha - \nu_0}{\lambda_m - \nu_0}, \quad (A6)$$

so that $0 \le x_0 \le 1$. It is a constant as far as the integration over $w$ is concerned.

In the thin-orbit limit we have $\lambda_1 = \lambda_m = \lambda_2$, i.e., $s = 0$, and hence

$$J_\lambda = 0,$$
$$\frac{\partial J_\lambda}{\partial s^2} = (\lambda_m + \alpha)\sqrt{\frac{(\lambda_m - \nu_0)U[\nu_0, \lambda_m, \lambda_m, \lambda_m]}{2(\lambda_m + \gamma)}}. \tag{A7}$$

Substitution in equation (3.4) for $c_g$ now immediately gives (3.5).

# APPENDIX B: APPROXIMATIONS AT LARGE RADII

The various third-order divided differences $U[\tau_0, \tau_1, \tau_2, \tau_3]$ that occur in the fundamental integral equation simplify when the potential $V$ becomes Keplerian $\propto -GM/\lambda^{-1/2}$ when $\lambda \to \infty$ (ZH, Hunter & de Zeeuw 1992). Here we need the case $\tau_0 \leq -\alpha$ and $-\alpha \ll \tau_1, \tau_2, \tau_3$. Upon substitution of the asymptotic behaviour $U(\lambda) \simeq -GM\lambda^{3/2}$ in the definition (2.9) we obtain

$$U[\nu, \lambda_1, \lambda_2, \lambda_3] \simeq \frac{GM}{(\sqrt{\lambda_1} + \sqrt{\lambda_2})(\sqrt{\lambda_1} + \sqrt{\lambda_3})(\sqrt{\lambda_2} + \sqrt{\lambda_3})},$$
$$U[\sigma, \nu, \lambda_1, \lambda_2] \simeq \frac{GM}{\sqrt{\lambda_1 \lambda_2}(\sqrt{\lambda_1} + \sqrt{\lambda_2})}. \tag{B1}$$

The function $U^*(\lambda, \nu; \nu_0, \lambda_1, \lambda_2)$ defined in equation (2.15) can therefore be approximated by

$$U^* \simeq \frac{(GM)^{3/2}}{4} \frac{(\sqrt{\lambda_1} + \sqrt{\lambda})^{1/2}(\sqrt{\lambda} + \sqrt{\lambda_2})^{1/2}}{\sqrt{\lambda_1 \lambda_2}(\sqrt{\lambda_1} + \sqrt{\lambda_2})^{7/2}}. \tag{B2}$$

In terms of the variables $s$ and $t$ this can be written as:

$$U^* \simeq \frac{(GM)^{3/2}}{2^{9/2}} \frac{(1+t)^{9/4}}{\lambda^{9/4}} L(s, t), \tag{B3}$$

where

$$L(s, t) = 2^{5/2} \frac{\left[\sqrt{1+t} + \sqrt{1-s}\right]^{1/2} \left[\sqrt{1+t} + \sqrt{1+s}\right]^{1/2}}{\sqrt{1-s^2}\left[\sqrt{1-s} + \sqrt{1+s}\right]^{7/2}}, \tag{B4}$$

so that $L(0, 0) = 1$ and $L(s, t)$ is even in $s$.

When $-\alpha \ll \lambda_1 \leq \lambda_2$, the integral (2.11) for the action $J_\lambda$ is elementary and independent of $\nu_0$. It is given by

$$J_\lambda \simeq \sqrt{2GM} \frac{(\lambda_2^{1/4} - \lambda_1^{1/4})^2}{(\sqrt{\lambda_1} + \sqrt{\lambda_2})^{1/2}}. \tag{B5}$$

Transformation to the variables $s$ and $t$ results in

$$J_\lambda \simeq \sqrt{2GM} \frac{\lambda^{1/4}}{(1+t)^{1/4}} \frac{\left[(1+s)^{1/4} - (1-s)^{1/4}\right]^2}{\left[\sqrt{1+s} + \sqrt{1-s}\right]^{1/2}}. \tag{B6}$$

Straightforward differentiation with respect to $s$ now gives

$$\frac{\partial J_\lambda}{\partial s^2} \simeq \frac{\sqrt{GM}}{4} \frac{\lambda^{1/4}}{(1+t)^{1/4}} h(s), \tag{B7}$$

where

$$h(s) = \frac{\sqrt{2}\left[4 + 2\sqrt{1-s^2} - (1-s^2)^{1/4}(\sqrt{1-s} + \sqrt{1+s})\right]}{(1-s^2)^{3/4}\left[\sqrt{1-s} + \sqrt{1+s}\right]^{5/2}}, \tag{B8}$$

so that $h$ is even in $s$. It is not difficult to show that $h(0) = 1$ and $h(s) > 1$ for $0 < s \leq 1$.

The normalization function $c_g(\nu_0, \lambda_m)$ can now be evaluated by substituting the above approximations in the definition (3.4). It becomes independent of $\nu_0$, and can be written as

$$c_g \simeq \frac{4}{C_g\sqrt{GM}} \frac{\lambda^{5/4}}{(1+t)^{5/4}}, \tag{B9}$$

where the constant $C_g$ is given by

$$C_g = \int_0^1 \mathrm{d}s^2 \, g_{\mathrm{sm}}(s) h(s). \tag{B10}$$

Since $h(s) \geq 1$, it follows that $C_g \geq 1$ for normalized $g_{\mathrm{sm}}$.

The intrinsic velocity moments are computed by inserting (3.18) as weight functions in the fundamental equation (3.17). With the help of equation (B1) we can approximate the velocities (3.18):

$$v_\lambda^2 = \frac{GM}{\sqrt{\lambda}} L_v^\lambda(s, t),$$
$$v_\nu^2 = \frac{GM}{\sqrt{\lambda}} L_v^\nu(s, t)(1 - u), \tag{B11}$$
$$v_\phi^2 = \frac{GM}{\sqrt{\lambda}} L_v^\phi(s, t) u,$$

where the $(s, t)$-dependent part has been separated:

$$L_v^\lambda = \frac{2(s^2 - t^2)}{(1+t)^{\frac{3}{2}}(\sqrt{1-s} + \sqrt{1+s})} \times$$
$$\frac{1}{(\sqrt{1+t} + \sqrt{1-s})(\sqrt{1+t} + \sqrt{1+s})};$$
$$L_v^\nu = \frac{2\left((1+t)^2 - s^2\right)}{\sqrt{1-s^2}(\sqrt{1-s} + \sqrt{1+s})\sqrt{1+t}}; \tag{B12}$$
$$L_v^\phi = \frac{2\sqrt{1-s^2}}{(\sqrt{1-s} + \sqrt{1+s})\sqrt{1+t}},$$

so that $L_v^\phi(0, 0) = L_v^\nu(0, 0) = 1$ and $L_v^\lambda(0, 0) = 0$.

# APPENDIX C: APPROXIMATIONS NEAR THE FOCAL CORNER

Near the focal corner in the $(\nu_0, \lambda_m)$-plane, where $\lambda_m = \nu_0 = -\alpha$, the function $U[\nu_0, \lambda_1, \lambda, \lambda_2]$ can be approximated by $U[-\alpha, -\alpha, -\alpha, -\alpha] = U'''(-\alpha) > 0$, so that it can be taken out of the integral for $J_\lambda$. It then follows that

$$\frac{\partial J_\lambda}{\partial s^2} \simeq \frac{(\lambda_m + \alpha)\sqrt{\lambda_m - \nu_0}}{\sqrt{2(\gamma - \alpha)}} \sqrt{U'''(-\alpha)} \, j(x_0, s), \tag{C1}$$

where

$$j(x_0, s) = \frac{1}{\pi} \int_{-1}^{1} \frac{\mathrm{d}w}{\sqrt{1-w^2}} \frac{\sqrt{1+(1-x_0)sw}}{1+sw}. \tag{C2}$$

The trigonometric substitution $w = \cos t$, followed by use of the integral tables of Byrd & Friedman (1971), shows that

$$j(x_0, s) = \frac{2}{\pi\sqrt{1+(1-x_0)s}} \left[(1-x_0)K(k) + \frac{x_0}{1+s}\Pi(\alpha^2, k)\right], \tag{C3}$$

where $K$ and $\Pi$ are the complete elliptic integrals of the first and third kind, respectively, with arguments given by

$$\alpha^2 = \frac{2s}{1+s}, \qquad k^2 = \frac{2(1-x_0)s}{1+(1-x_0)s}. \tag{C4}$$

In the thin-orbit limit $j(0,0) = 1$, so that expression (C1) reduces to (A7) evaluated at $\lambda_m = \nu_0 = -\alpha$. Two special cases of interest are

$$j(0,s) = \frac{2}{\pi\sqrt{1+s}} K(k) = {}_2F_1(\tfrac{1}{4}, \tfrac{3}{4}; 1; s^2),$$
$$j(1,s) = \frac{1}{\sqrt{1-s^2}}. \tag{C5}$$

Here we have used formulae 8.114 and 9.134.1 of Gradshteyn & Ryzhik (1980, hereafter GR), to write $j(0,s)$ explicitly as a (hypergeometric) function of $s^2$.

The function $c_g$ can now be approximated as

$$c_g(\nu_0, \lambda_m) \simeq \frac{\sqrt{2(\gamma - \alpha)}}{\sqrt{U'''(-\alpha)}} \frac{1}{J_g(x_0)}, \tag{C6}$$

with

$$J_g(x_0) = \int_0^1 \mathrm{d}s^2\, g_{\mathrm{sm}}(s) j(x_0, s). \tag{C7}$$

This shows that $c_g(\nu_0, \lambda_m)$ has radial behaviour near the focal corner: its value depends on the direction along which the focal corner is approached. The function $J_g(x_0)$ equals 1 for all $x_0$ when $g_{\mathrm{sm}} = \delta(s^2)$, and it varies slowly with $x_0$ for sharply peaked $g_{\mathrm{sm}}$.

For $s_{\max} < 1$ we can express $J_g(x_0)$ in terms of the moments (3.7) of $g_{\mathrm{sm}}$ by expanding $j(x_0, s)$ in powers of $s^2$. We give the result for the two special cases $x = 0$ and $x = 1$:

$$J_g(0) = \frac{1}{\pi\sqrt{2}} \sum_{k=0}^{\infty} \frac{\Gamma(k+\tfrac{1}{4})\Gamma(k+\tfrac{3}{4})}{k!} \langle s^{2k} g_{\mathrm{sm}} \rangle,$$
$$J_g(1) = \frac{1}{\sqrt{\pi}} \sum_{k=0}^{\infty} \Gamma(k+\tfrac{1}{2}) \langle s^{2k} g_{\mathrm{sm}} \rangle. \tag{C8}$$

$J_g(0)$ and $J_g(1)$ can be evaluated explicitly for the functions $g_{\mathrm{sm}}$ defined in equation (3.6). Use of GR 7.512.11 gives

$$J_g(0) = {}_2F_1(\tfrac{1}{4}, \tfrac{3}{4}; 2+q; s_{\max}^2),$$
$$J_g(1) = {}_2F_1(\tfrac{1}{2}, 1; 2+q; s_{\max}^2). \tag{C9}$$

These hypergeometric functions equal one for $s_{\max} = 1$, and are well-behaved for $q \geq 0$ and $0 \leq s_{\max} \leq 1$. For $q = 0$ we can write

$$J_g(0) = \frac{8\sqrt{1+s_{\max}}}{3\pi s_{\max}^2}\big[E(k) - (1-s_{\max})K(k)\big],$$
$$J_g(1) = \frac{2}{1+\sqrt{1-s_{\max}^2}}, \tag{C10}$$

where $E$ is the complete elliptic integral of the second kind, and $k^2 = 2s_{\max}/(1+s_{\max})$. When $q = 0$ we have

$$J_g(0) = \frac{\Gamma(2+q)\Gamma(1+q)}{\Gamma(\tfrac{7}{4}+q)\Gamma(\tfrac{5}{4}+q)},$$
$$J_g(1) = \frac{1+q}{\tfrac{1}{2}+q}. \tag{C11}$$

These expressions are valid for $s_{\max} = 1$ as well, in which case $J_g(0) = 16/3\pi\sqrt{2}$ and $J_g(1) = 2$.

## APPENDIX D: THE INTEGRAL (4.22)

When $x = 0$ we have $x_0 = 0$, and equation (4.22) reduces to

$$F_g(0) = \frac{1}{\pi^2 J_g(0)} \int_0^1 \frac{\mathrm{d}u}{\sqrt{u(1-u)}} \int_0^{s_{\max}^2} \mathrm{d}s^2 \int_{-s}^s \frac{\mathrm{d}t\, g_{\mathrm{sm}}(s)}{(1+t)^{3/2}\sqrt{s^2-t^2}}. \tag{D1}$$

The $u$-integration is a factor, and is easily evaluated as $\pi$ upon the trigonometric substitution $u = \cos^2 z$. The $t$-integral equals $2E(k)(1+s)^{-1/2}(1-s)$, where $E(k)$ is the complete elliptic integral of the second kind and $k^2 = 2s/(1+s)$. We express it as a hypergeometric function by means of formula GR 3.133.15, and use GR 9.134.1 to write it as a function of $s^2$. The result is

$$F_g(0) = \frac{1}{J_g(0)} \int_0^{s_{\max}^2} \frac{\mathrm{d}s^2\, g_{\mathrm{sm}}(s)}{1-s^2}\, {}_2F_1(-\tfrac{1}{4}, \tfrac{1}{4}; 1; s^2). \tag{D2}$$

The case $x = 1$ can be done in a similar way. Now $x_0 = 1$, and we have

$$F_g(1) = \frac{2}{\pi^2 J_g(1)} \int_0^1 \frac{\sqrt{u}\,\mathrm{d}u}{\sqrt{1-u}} \int_0^{s_{\max}^2} \mathrm{d}s^2 \int_{-s}^s \frac{g_{\mathrm{sm}}(s)\mathrm{d}t}{(1+t)\sqrt{1-s^2}\sqrt{s^2-t^2}}. \tag{D3}$$

The $u$-integration now gives $\pi/2$, while the $t$-integral follows from GR 3.163.1 after the substitution $sx = \cos z$, and results in $\pi(1-s^2)^{-1/2}$. This leaves

$$F_g(1) = \frac{1}{J_g(1)} \int_0^{s_{\max}^2} \frac{\mathrm{d}s^2\, g_{\mathrm{sm}}(s)}{1-s^2}. \tag{D4}$$

For $s_{\max} < 1$ we can express these integrals in terms of the moments $\langle s^{2k} g_{\mathrm{sm}} \rangle$ of $g_{\mathrm{sm}}$ defined in equation (3.7) by expanding the integrand in powers of $s^2$. This gives

$$F_g(0) = \frac{1}{J_g(0)} \frac{4}{\pi\sqrt{2}} \sum_{k=0}^{\infty} \frac{\Gamma(k+\tfrac{3}{4})\Gamma(k+\tfrac{5}{4})}{k!} \langle s^{2k} g_{\mathrm{sm}} \rangle,$$
$$F_g(1) = \frac{1}{J_g(1)} \sum_{k=0}^{\infty} \langle s^{2k} g_{\mathrm{sm}} \rangle. \tag{D5}$$

The integrals (D2) and (D4) can be evaluated explicitly for the specific choice (3.6) for the function $g_{\mathrm{sm}}(s)$. Use of GR 3.197.3 gives

$$F_g(0) = \frac{1}{J_g(0)}\, {}_2F_1(\tfrac{3}{4}, \tfrac{5}{4}; 2+q; s_{\max}^2),$$
$$F_g(1) = \frac{1}{J_g(1)}\, {}_2F_1(1, 1; 2+q; s_{\max}^2), \tag{D6}$$

with $J_g(0)$ and $J_g(1)$ given in equation (C9). These hypergeometric functions equal one when $s_{\max} = 0$, and are well-behaved for $q > -1$ and $0 \leq s_{\max} < 1$. For $s_{\max} = 1$ we obtain

$$F_g(0) = \frac{1}{J_g(0)} \frac{\Gamma(2+q)\Gamma(q)}{\Gamma(\tfrac{3}{4}+q)\Gamma(\tfrac{5}{4}+q)},$$
$$F_g(1) = \frac{1}{J_g(1)} \frac{1+q}{q}, \qquad (q > 0). \tag{D7}$$

When $q = 0$ we find

$$F_g(0) = \frac{8\left[K(k) - (1+s_{\max})E(k)\right]}{\pi s_{\max}^2 \sqrt{1+s_{\max}} J_g(0)},$$
$$F_g(1) = \frac{-1}{s_{\max}^2} \ln(1 - s_{\max}^2),$$
(D8)

with $k^2 = 2s_{\max}/(1+s_{\max})$. It follows that $F_g(0)$ and $F_g(1)$ diverge logarithmically when $s_{\max} \to 1$, but their ratio approaches $3/2$.